\documentclass[10pt,journal,compsoc]{IEEEtran}
\ifCLASSOPTIONcompsoc
  \usepackage[nocompress]{cite}
\else
  \usepackage{cite}
\fi
\ifCLASSINFOpdf
\else
\fi
\usepackage{graphicx}

\usepackage{url}
\makeatletter
\def\Url@twoslashes{\mathchar`\/\@ifnextchar/{\kern-.2em}{}}
\g@addto@macro\UrlSpecials{\do\/{\Url@twoslashes}}
\makeatother
\usepackage{hyperref}
\usepackage{multirow}
\usepackage{makecell}
\begin{document}
\title{DevOps Team Structures: Characterization and Implications}

\author{Daniel López-Fernández, %~\IEEEmembership{Member,~IEEE,}
        Jessica Díaz, %~\IEEEmembership{Fellow,~OSA,}
        Javier García, %~\IEEEmembership{Fellow,~OSA,}
        Jorge Pérez, %~\IEEEmembership{Fellow,~OSA,}
        and~Ángel González-Prieto %~\IEEEmembership{Life~Fellow,~IEEE}% 
        % stops a space
\IEEEcompsocitemizethanks{\IEEEcompsocthanksitem D. López-Fernández and J. Díaz and J. García and J. Pérez and A. González-Prieto were with the Universidad Politécnica de Madrid (ETSI de Sistemas Inform\'aticos), 28031 Madrid, Spain.\protect\\
E-mail: \{daniel.lopez, yesica.diaz, jgarciam, jorge.perez, angel.gonzalez.prieto\}@upm.es
%\IEEEcompsocthanksitem J. Doe and J. Doe are with Anonymous University.
}% <-this % stops a space
%\thanks{Manuscript received January 19, 2020; revised August 26, 2015.}
}


\markboth{Journal of \LaTeX\ Class Files,~Vol.~14, No.~8, August~2015}%
{Shell \MakeLowercase{\textit{et al.}}: Bare Advanced Demo of IEEEtran.cls for IEEE Computer Society Journals}



\IEEEtitleabstractindextext{%
\begin{abstract}

\textbf{Context}: DevOps can be defined as a cultural movement to improve and accelerate the delivery of business value by making the collaboration between development and operations effective. 
\textbf{Objective}:  This paper aims to help practitioners and researchers to better understand the organizational structure and characteristics of teams adopting DevOps. 
\textbf{Method}: We conducted an exploratory study by leveraging in depth, semi-structured interviews to relevant stakeholders of 31 multinational software-intensive companies, together with industrial workshops and observations at organizations' facilities that supported triangulation. We used Grounded Theory as qualitative research method to explore the structure and characteristics of teams, and statistical analysis to discover their implications in software delivery performance.
\textbf{Results}: We describe a taxonomy of team structure patterns that shows emerging, stable and consolidated product teams that are classified according to six variables, such as collaboration frequency, product ownership sharing, autonomy, among others, as well as their implications on software delivery performance.  
%which may range from teams with a sporadic collaboration between developers and operators to daily collaboration of autonomous and cross-functional teams. 
These teams are often supported by horizontal teams (DevOps platform teams, Centers of Excellence, and chapters) that provide them with platform technical capability, mentoring and evangelization, and even temporarily facilitate human resources. 
%(i) Interdepartmental Dev \& Ops collaboration; (ii) Interdepartmental Dev-Ops teams, both supported by DevOps Platform teams; (iii) Boosted cross-functional DevOps product teams, supported by Immersive DevOps Centers of Excellence that help them to achieve the \textit{you build it, you run it} capability, in addition to providing tools. Full cross-functional DevOps team, evangelized and mentored by DevOps Centers of Excellence. 
\textbf{Conclusion}: This study aims to strengthen evidence and support practitioners in making better informed about organizational team structures by analyzing their main characteristics and implications in software delivery performance.
\end{abstract}

\begin{IEEEkeywords}
DevOps, Team Structures, Grounded Theory.
\end{IEEEkeywords}}

\maketitle

% is not selected <OR> if conference mode is selected - because compsoc
% conference papers position the abstract like regular (non-compsoc)
\IEEEdisplaynontitleabstractindextext

% \fi
\IEEEpeerreviewmaketitle

\ifCLASSOPTIONcompsoc
\IEEEraisesectionheading{\section{Introduction}\label{sec:introduction}}
\else
\section{Introduction}
\label{sec:introduction}
\fi


\IEEEPARstart{D}{evOps} is an organizational transformation that had its origin at the 2008 Agile Conference in Toronto, where P. Debois highlighted the need to resolve the conflict between development and operations teams when they had to collaborate to provide quick response time to customer demands \cite{Debois:2008}. Later, at the O’Reilly Velocity Conference, two Flickr employees delivered a seminal talk known as \textit{``10+ Deploys per Day: Dev and Ops Cooperation at Flickr''}, which can be considered the starting point to extend agility beyond development \cite{Allspaw:2009}. 

To enlighten the problems and drivers that are currently moving companies to adopt DevOps, we conducted an empirical research \cite{ese} that shows the most important concerns when adopting DevOps, such as the excessive time for releasing,  problems when releasing new versions, digital transformation drivers (e.g., market trends, technological obsolescence), and lack of standardization and automation. Companies with these problems shift to DevOps due to the associated benefits that DevOps claim, such as an increase on software delivery performance, process productivity, and team effectiveness, which results in faster time-to-market, better software quality, and greater alignment of developers and operators with business goals and customer focus \cite{ese}. %In short, DevOps promotes fast and frequent delivery of new and changing features while ensuring the quality and non-disruption of the production environment and customers \cite{Lwakatare:2016}.

Such has been the spread of DevOps worldwide that, on many occasions, has been misinterpreted and referred to as a set of tools for automating software process. %, such as building, testing, deployment, infrastructure creation, etc. 
Nevertheless, beyond all that, DevOps is a cultural movement that aims for collaboration among all stakeholders involved in the development, deployment and operation of software to deliver a high-quality product or service in the shortest possible time \cite{LuzPinto:2019}. DevOps promotes a collaborative culture based on \textit{product teams} that share business goals and an end-to-end product vision. Thus, the DevOps culture largely determines how business people, developers and operators are organized, and vice-versa, how business people, developers and operators are organized determines how DevOps culture is adopted and widespread across organizations. However, how to organize and structure these teams is still a challenge. %Is it as simple as putting developers and operators to work together? %There exist great differences in the way in which DevOps promotes work and the traditional way in which most software companies have been working for decades. Because of this, 
Companies need evidence about successful---or not such successful but less intensive-resource---organizational team structures that work on a daily basis. While everyone agree that effective teams enable organizations to continuously deliver business value, practitioners and researchers \cite{Smeds:2015,Lwakatare:2016,Senapathi:2018} also agree on considering the structure of teams and their interactions as main challenges that companies face during DevOps transformation processes. Hence, the research questions we can extract are: \vspace{0.1cm}

{\setlength{\parindent}{0pt}\textbf{RQ1} \textit{How do real-world organizations structure themselves to instilling a DevOps culture into their organization?}}

{\setlength{\parindent}{0pt} \textbf{RQ2} \textit{What implications do team structure patterns have on software delivery performance? }}

To address RQ1, we engaged in theory building based on empirical data \cite{stol:2016,sjoberg:2008,ralph:2019} to understand \textit{organizational  team structures}, their communication and collaboration pathways, as key drivers for adopting DevOps culture and practices. %, rapid software delivery, and continuous experimentation. 
To address RQ2, we statistically analyzed whether different team structures could impact on well-known software delivery performance indicators, such as lead time, mean time to recovery, and delivery frequency. 

%To do so, we focused on understanding the team structures, cultural and sharing practices, and communication and collaboration pathways by studying multiple cases of “representative” companies.  

Methodologically, we applied Grounded Theory \cite{glaser:1967, charmaz:2014}, a method well-suited for generating theories. We aim to generate a theory in the form of \textbf{a taxonomy of team structure patterns} in the context of 31 organizations that have been adopting DevOps for at least two years (i.e., how teams are organized, how people collaborate and break silos, how autonomous they are, etc.). In this regard, in two prior conference papers \cite{Diaz:2018, Diaz:2019}, we described the research methodology of a global exploratory study on practicing DevOps and a preliminary analysis of team structures based on an initial set of companies. These preliminary results were discussed with the research and industrial community, and improved and refined later with more collected data until we got saturation. Thus, the paper here presented differs from this previous work in building a full theory on DevOps team patterns by conducting a GT study and statistical analysis to answer the research questions. Additionally, we conducted a clustering analysis to prove whether the taxonomy we obtained with qualitative analysis is similar to the clusters that were obtained using a K-means algorithm. 

The structure of the paper is as follows: Section~\ref{sec:methodology} describes the research methodology. Section~\ref{sec:theory} reports the emerging theory as a conceptual framework and Section~\ref{sec:model} describes a taxonomy of DevOps team structure patterns (RQ1). Section~\ref{sec:statistical} describes patterns implications on software delivery performance (RQ2) together clustering analysis to assess RQ1.  Section~\ref{sec:validity} assesses the validity and reliability of these outcomes. Section~\ref{sec:relatedwork} describes related work. Finally, conclusions and further work are presented in Section~\ref{sec:conclusion}.


\section{Research Methodology} %Research Design
\label{sec:methodology}

The research presented here is part of a larger investigation on practicing DevOps we started in 2017 \cite{Diaz:2018, Diaz:2019, ese}. This research is mainly based on the \textit{constructivism} model as underlying philosophy \cite{Easterbrook:2008}. Constructivism or interpretivism states that scientific knowledge cannot be separated from its human context, and a phenomenon can be fully understood by considering the perspectives and the context of the involved participants \cite{Easterbrook:2008}. Therefore, the most suitable methods to support this approach are those collecting rich qualitative data, from which theories (tied to the context under study) may emerge. 

According to this, we approached diverse qualitative data through a multiple case study. \textit{“A case study is an empirical inquiry that investigates a contemporary phenomenon within its real-life context”}, in which \textit{“the boundaries between phenomenon and context may not be clearly evident”}, and \textit{“relies on multiple sources of evidence, with data needing to converge in a triangulating fashion”} \cite{Yin:2018}. In this regard, we have been studying the context of DevOps adoption and practice in +30 multinational software-intensive companies from November 2017 to nowadays, mainly by leveraging in-depth, semi-structured interviews with relevant stakeholders of these companies, together with industrial workshops and observations at organizations’ facilities that supported triangulation and allowed us to complete data omissions in the interviews when necessary. Interviews are a common method used for collecting data in software field studies \cite{Lethbridge:2005} and, together observations, provided us with the necessary data to answer the research questions of this study. %[Our case study is \textit{exploratory} in the sense it seeks what is happening in a phenomenon while also seeks new insights and generating ideas for new research \cite{Wohlin:2012} and \textit{interpretivist} in  the sense it studies a social phenomenon and requires different empirical methods and assumptions than studying physical phenomena \cite{ralph:2019}.] %“Interpretivist case studies generate theories, taxonomies or other concepts from data, rather than testing a priori theories or artifacts” \cite{ralph:2019}. 

In a previous work \cite{ese} we addressed two research questions related to the \textbf{problems} that companies try to solve by implementing DevOps and the \textbf{results}  they try to achieve. In that research, we used \textit{thematic analysis} \cite{Cruzes:2011}, and specifically an \textit{integrated approach} based on a deductive approach for creating themes that we retrieved from literature, and an inductive approach for creating codes. Using these two approaches, some specific fragments of the interviews were coded by two researchers in an iterative process until the inter-coder agreement reached an acceptable level of reliability and, finally, a theory was built through a synthesis process. 

At the same time as we were interviewing more and more organizations and knowing why companies are instilling a DevOps culture in their organizations, we began to glimpse that organizations talked about different structures of their teams to try to address the main principles of DevOps culture, align business objectives with development and operations teams, increase collaboration and communication between these teams, and accelerate value delivery to customers in a continuous improvement and experimentation process. At that time (2018), little work had empirically analyzed team structures in companies adopting DevOps \cite{shahin:2017} or reported team topologies based on success cases \cite{skelton:2013}. 

In this paper, we decided to use the \textbf{grounded theory} (GT) method and started this study on DevOps team structure. GT is a technique for iteratively developing theory  from qualitative data \cite{glaser:1967} that encourages a deep immersion in the data \cite{ralph:2019}. \textit{“In grounded theory, initial analysis of the data begins without any preconceived categories. As interesting patterns emerge, the researcher repeatedly compares these with existing data, and collects more data to support or refute the emerging theory”} \cite{Easterbrook:2008}. Salleh et al. \cite{salleh:2018} assert that GT can be adopted as a wrapper around other empirical methods such as case studies where GT analysis and theory formulation procedures are applied to data collected from case studies, as in \cite{Bick:2018,Halaweh:2008}. Thus, GT is adequate for our purposes.

GT is increasingly being used to study software engineering research topics \cite{stol:2016}, such as to characterize scenarios under a personal perspective of those engaged in a culture, which is the topic here addressed. Most SE studies select the classic or Glaserian GT variant \cite{glaser:1967}, however, according to our philosophical stance (epistemological and ontological positions), we selected  the Charmaz's \textit{constructivist GT} variant \cite{charmaz:2014}. Constructivist GT emphasizes that understanding and acknowledges data, interpretations, and resulting theory depend on the researcher’s view \cite{stol:2016}. As Stol et al. \cite{stol:2016} assert, in practice, such ontological and epistemological differences are rarely apparent in generated theories, however, the existence of an upfront research question, the role of the literature, the terminology, and the order of practices differ from one variant to another. 

As GT literature states, our study involved simultaneous data collection and analysis, i.e., coding and memoing \cite{Saldana2012}, constant comparison and cross-case analysis \cite{seaman}, and theory development \cite{sjoberg:2008}. The following sections describe data collection, data analysis, and validity procedures.

%State the research area or research question—either your initial question, the question that emerged during your study, or preferably both.
%State the duration of the study

\subsection{Data collection}
%Describe the context of the study (e.g. the kind of organization, who is involved, what kind of software is being developed).
GT involves iteratively performing interleaved rounds of qualitative data collection and analysis to lead to theory (e.g., concepts, categories, patterns) \cite{ralph:2020}. The selection of participants is also iterative and can be considered a combination of \textit{theoretical sampling}, in the sense that we chose which data to collect based on the concepts or categories that were relevant for the emerging theory, i.e., organizations that have been adopting DevOps for at least two years; \textit{maximum variation sampling}, in the sense that we tried to choose highly diverse people and organizations in our sample, strengthening the transferability of our theory; and \textit{convenience sampling} as we are restricted to organizations and relevant stakeholders to which we had access. This iterative process ended when \textit{theoretical saturation} became apparent, which means that processing new data did not impacted the theory elaborated until that moment%We reached saturation with 28 companies when 
---the last few participants provided more evidence and examples but no new concepts or codes. 
In fact, we included some more companies until we could be sure that new participants would not add new knowledge.

Table~\ref{tab:Subjects} lists the organizations involved in the study, its ID, scope (international or national), size\footnote{Spanish Law 5/2015 indicates that a \textit{micro enterprise} is one that has less than ten workers and an annual turnover of less than two million euros or a total asset of less than two million euros; a \textit{small company} is one that has a maximum of 49 workers and a turnover or total assets of less than ten million euros;  \textit{medium-sized companies} are those with less than 250 workers and a turnover of less than fifty million euros or an asset of less than 43 million euros; and \textit{large companies} are those that exceed these parameters.}, business core and organization age\footnote{Consulting firms denoted by * provided the data of one of their clients.}. In many cases, more than one participant participated in an interview. Indeed, a total of 46 people were enrolled in the interviews. Table~\ref{tab:Interviewees} provides anonymized information about the position and IT experience of the interviewees. We interviewed key stakeholders, such as CEOs, CIOs, DevOps platform leaders, product leaders, developers, and infrastructure managers (see Table \ref{tab:Interviewees}). These interviewees had all the necessary information to adequately answer the question posed. %Moreover, in some cases we contacted with the interviewees several times until all questions or some misinterpretation were solved. 

\renewcommand{\arraystretch}{1.1}

\begin{table}[t]
\centering
\caption{Subject description}
\small
\begin{tabular}{p{0.5cm} p{1.7cm} p{1.1cm} p{2cm} p{1.5cm}}
\hline
\textbf{Id} & \textbf{Scope} & \textbf{Size} &\textbf{Business} &\textbf{Creation date} \\ 
\hline 
01 & International & Medium & Retail &  2000-2010 \\  
02 & National & Large & Retail & $<$2000 \\ 
03 & International & Medium & Software & 2000-2010 \\
04 & National & Large & Telecom & $<$2000 \\ 
05 & National & Large & Public Utility & $<$2000 \\ 
06 & International & Large & *Banking & $<$2000 \\ %\newline $\Rightarrow$ Banking
07 & National & Large & Educational & $<$2000  \\ 
08 & National & Large & *N/A  & $<$2000 \\ %$\Rightarrow$~N/A
09 & International & Large & FinTech & 2000-2010 \\ 
10 & National & Medium & *Logistic & $<$2000 \\ %$\Rightarrow$~Logistic
11 & International & Medium & Retail & $<$2000 \\ 
12 & International & Large & Logistic & $<$2000\\ 
13 & International & Large & Retail & $<$2000  \\ 
14 & International & Large & Telecom & $<$2000 \\ 
15 & National & Large & *Telecom & $<$2000 \\ %$\Rightarrow$~Telecom
16 & National & Large & *Banking & $<$2000 \\ %$\Rightarrow$~Banking
17 & International & Large & Telecom & 2000-2010 \\ 
18 & International & Large & Real estate & $<$2000 \\ 
19 & International & Large & *Banking & $<$ 2000 \\ %$\Rightarrow$~Banking
20 & National & Large & Insurance & $<$2000 \\ 
21 & National & Large & *Marketing & 2000-2010 \\ %$\Rightarrow$~Marketing
22 & International & Small & *Retail & $>$2010 \\ %$\Rightarrow$~Retail
23 & International & Large & Telecom & $<$2000 \\ 
24 & International & Large &  *N/A & $<$2000 \\ %$\Rightarrow$~N/A
25 & International & Large & *Telecom & $<$2000 \\ %$\Rightarrow$~Telecom
26 & National & Large & Banking & $<$ 2000 \\ 
27 & International & Large & *N/A & $<$2000 \\ %$\Rightarrow$~N/A
28 & International & Large & Marketplace & 2000-2010 \\ 
29 & International & Large & Retail & $<$2000 \\ 
30 & International & Large & *Banking & 2010 \\ %$\Rightarrow$~Banking
31 & International & Large & Energy & $<$2000 \\ 
\hline
\end{tabular}
\label{tab:Subjects}
\end{table}

\begin{table}[t]
\centering
\caption{(Anonymized) Description of Interviewees}
\small
\begin{tabular}{p{3.5cm} p{1.2cm} }
\hline
\textbf{Position} & \textbf{Number}\\ 
\hline 
Executive manager & 11 \\  
Service manager & 3 \\ 
Infrastructure manager & 8 \\
Project manager & 11 \\ 
Consultant & 5 \\ 
Developer & 8 \\ 
\hline 
\textbf{Experience (years)} & \textbf{Number}\\
\hline 
$+$20 & 17\\ 
16-20 & 12\\
11-15 & 11\\ 
5-10 & 6\\ 
\hline
\end{tabular}
\label{tab:Interviewees}
\end{table}
\renewcommand{\arraystretch}{1.0}

%*The researcher got to this display (conceptual framework) after spending some time observing and interviewing*
The interview protocol and its script are described in \cite{ese} and \url{https://blogs.upm.es/devopsinpractice/}. It is necessary to highlight that we only focused on a set of specific fragments of the collected data from interviews for this particular research, together with three software delivery performance indicators \cite{dora:2018}: lead time, deployment frequency and mean time to recovery. 

Additionally, we had the opportunity to attend working sessions in the organizations’ facilities. In these sessions we observed IT departments in their daily work. In this way, we observed first-hand product poly-skilled and cross-functional teams, development and infrastructure teams, how they collaborated (e.g., they showed us some content from Confluence, Jira, Slack) or how they did not collaborate. %In some cases, they even showed us information radiators, where we collected data related to problems (e.g., delays in releases, system downtime, open issues, fail rates of pipeline executions). 
On other occasions, we organized industrial workshops with some participant organizations at the university campus\footnote{\url{http://bit.ly/2ky00LQ}, last accessed January 2020.}. Therefore, we managed to involve the participants in the study beyond the interviews. Finally, we can mention that some participants showed us different snapshots of a DevOps maturity framework, from which we could extract organizational changes over teams they addressed. We obtained many observations of these visits (gathered in a \textit{research diary}) that we analyzed together with the interviews, mainly to triangulate data and complete data omissions. 

The outputs of this phase are the transcriptions of the interviews and observations.

\subsection{Data analysis}

%Who collected and analyzed the data? Was it an individual researcher or research team? If a team, who did what? How was this coordinated? Describe your coding, memoing and sorting with examples.
We used the guidelines about how to conduct a constructivist GT reported by Stol et al. \cite{stol:2016}, which organize a GT investigation in three phases:

\textbf{Initial coding}: This phase is mainly based on the principles of \textit{coding},  \textit{constant comparison} and \textit{cross-case analysis} \cite{seaman}. The inputs of this phase are the transcripts of the interviews. These transcripts are coded, which means that parts of the text can be given a code representing a certain \textit{construct}, by examining data word-by-word, line-by-line or incident-by-incident to make sense of the text without injecting the researcher’s assumptions, biases, or motivations \cite{stol:2016}. One code is usually assigned to many pieces of text, and one piece of text can be assigned more than one code. Codes can form categories, i.e., a hierarchy of codes and sub-codes. While coders analyze the data, they write memos, i.e., notes about ideas or concepts potentially relevant to the research. As more interviews are analyzed, coders constantly compares the data, memos and codes of the same interview and other interviews (cross-case), which forces the researchers to go back and forth.  

\textbf{Focused coding}: In this phase, the researchers take the codes resulting from the initial coding and select categories from the most frequent or important codes, i.e. \textit{variables}, and use them to categorize the data \cite{stol:2016}. Memos elaborate categories, describe properties and relationships between categories, and identify gaps \cite{charmaz:2014}. The researchers attempt to move beyond superficial categories to develop a cohesive theory of the studied phenomenon \cite{stol:2016}. %; does not require a single core category or variable.

\textbf{Theoretical coding}:  In this phase, the researchers are involved in a continuous process of oscillating between the memos, categories, and the emerging theory to specify the relationship between categories to integrate them into a cohesive theory \cite{stol:2016}. When no new categories or relationships between them emerge, theoretical saturation is reached, i.e., new data is no longer generating reinterpretations of the theory. 

These three phases were iteratively carried out as new data (interviews and observations) were collected. The outputs of these phases are the coding, the memos, the categories, and the resulting theory, all of them managed through the tool Atlas.ti v8 \cite{Atlas:2019}. %The two first phases were performed by the two first co-authors, while the other three co-authors participated in the theoretical coding. 
To support our findings, we included excerpts from these interviews in our chain of evidence. 

\subsection{Validity procedure}
\label{subsec:validityprocedures}

Criteria for judging the quality of research designs are key to establish the validity, i.e., the accuracy of the findings, and the reliability, i.e. the consistency of procedures and researcher’s approach, of most empirical research \cite{Yin:2018,Creswell:2017}. There are different ways to classify threats to validity and reliability in the literature \cite{Wohlin:2012}. Glaser \cite{glaser:1967}, Strauss and Corbin \cite{strauss:1990} and Charmaz \cite{charmaz:2014} describe inconsistent quality criteria for GT. We considered the quality criteria defined by Charmaz together the quality criteria defined by Lincoln and Guba’s \cite{Lincoln:1985} for interpretative qualitative research in general, which have been incorporated in the ACM SIGSOFT Empirical standards \cite{ralph:2020}.

Taking all these references as a basis, quality criteria can be described as follows: \textit{credibility}, i.e., the extent to which conclusions are supported by rich, multivocal evidence; \textit{resonance}, i.e., the extent to which a study’s conclusions make sense to (i.e., resonate with) participants; \textit{usefulness}, i.e., the extent to which a study provides actionable recommendations to researchers, practitioners or educators and the degree to which results extend our cumulative knowledge; \textit{transferability}, which shows whether the findings could plausibly apply to other situations; \textit{dependability}, which shows that the research process is systematic and well documented and can be traced; and \textit{conformability}, which assesses whether the findings emerge from the data collected from cases and not from preconceptions. 

To satisfy most of these criteria, we followed the strategies pointed out by Creswell \cite{Creswell:2017} to improve the validity of qualitative research, as follows. 

{\setlength{\parindent}{0pt} \textit{1. Triangulation}. As said by N. Denzin \cite{Norman}, “\textit{the greater the triangulation, the greater the confidence in the observed findings}”. In this study we established to conduct data triangulation, i.e., data collected for different times, locations, populations, etc., and method triangulation, i.e., using data collected by different methods.}

{\setlength{\parindent}{0pt} \textit{2. Member checking} enables to determine the accuracy of the findings \cite{Creswell:2017}. We established a milestone to show our ongoing work and receive feedback from a subset of participants to determine whether they feel that preliminary results are accurate.} %credibility

{\setlength{\parindent}{0pt} \textit{3. Rich description} enables to convey the findings \cite{Creswell:2017}. We described the context of the involved organizations and teams as much as confidentiality issues allow.} %This may enable to consider \textit{confounding factors} that may have any effect on the outcomes.}

{\setlength{\parindent}{0pt} \textit{4. Clarify bias} enables to reflect how researchers interpretation of the findings is shaped by their background \cite{Creswell:2017}. We tried to avoid bias and negative impacts on some participants or stakeholders visible (i.e., reflexivity \cite{ralph:2020}) through double-checks of transcriptions and coding. 
%so that researchers should , i.e. their potential biases and interactions with the team, organization or community, especially possible negative impacts on some participants or stakeholders (reflexivity) \cite{ralph:2020}.
} 

{\setlength{\parindent}{0pt} \textit{5. Report discrepant information} so that all the results are presented and analyzed, regardless of their implications for our initial interests. Prolonged contact with participants, the duration of the interviews, and the subsequent communication with them, allowed us to fully understand their perspectives to mitigate some information that contradicts the general perspective of the theory.}

{\setlength{\parindent}{0pt} \textit{6. Spend prolonged time in the field}. The researchers developed an in-depth understanding of the phenomenon under study to convey details about the site and the people that lends credibility to the narrative account.} %credibility

{\setlength{\parindent}{0pt} \textit{7. Use an external auditor to review the project} provides an objective assessment of the project throughout the research process and study conclusions \cite{Creswell:2017}. We established a milestone to publish our ongoing work and receive feedback from researchers and practitioners in an international conference in the middle of the study.}

Section~\ref{sec:validity} describes the threats to validity of this study and how the abovementioned strategies were conducted to try to mitigate them. 

%These criteria were be addressed during all phases of the GT study.

%How does the literature inform, support or refute your analysis and results?
%How might your own biases, preconceptions, background and beliefs affect your analysis?
%Did you conduct a reliability check; i.e., have your analysis reviewed by someone else. If so, who, how, what did they find and what changes resulted? Describe their expertise

%%%%%%%%%%%%%%%%%%%%%%%%%%%%%%%%%%%%%%%%%%%%%%%%%%%%%%%%%%%%%%%%%%%%%%%%%%%%%%%%%%%%%%%%%%%%%%%%%%%%
%%%%%%%%%%%%%%%%%%%%%%%%%%%%%%%%%%%%%%%%%%%%%%%%%%%%%%%%%%%%%%%%%%%%%%%%%%%%%%%%%%%%%%%%%%%%%%%%%%%%

\section{A theory on DevOps Team Structures}
\label{sec:theory}

This section describes the conduction of the GT study according to the procedures abovementioned. Data collection (i.e., selection of participants, interviews, and transcriptions) and data analysis (i.e., initial coding, focused coding, and theoretical coding) were simultaneously performed. %Hence, the interviews were analyzed by examining each text excerpt (quotation) and by assigning them one or more codes representing a certain construct. While the coders were analyzing the interviews, they also took notes about ideas or concepts potentially relevant to the research (i.e., memoing). 
As the study progressed, categories from codes and memos were built through constant comparison and cross-case analysis; then, key variables from these categories were selected; finally, a cohesive theory on DevOps team structures was built. The following subsections describe these three iterative phases together with raw data quotations of the interviews (see boxes below and the ID of the participant) to make explicit the chain of evidence.

\subsection{Categories and Concepts}

Here we detail our understanding of the core category, which we refer to as \textbf{DevOps organizational team structures}, and subcategories: \textbf{product teams}, \textbf{horizontal (cross) teams}, \textbf{silos}, \textbf{culture values}, and \textbf{best practices}. Subcategories and codes are highlighted throughout the text using italics along with excerpts of the interviews to make explicit the chain of evidence.

\textbf{Product team} category emerges as a result of a set of codes that characterize these teams: \textit{collaboration}, \textit{product ownership sharing}, \textit{end-to-end product vision}, \textit{cross-functionality} (sometimes used as synonym of \textit{multidisciplinary} or \textit{poly-skilled} teams), \textit{self-organization} and \textit{autonomy}. Product teams are usually small (\textit{Amazon's two pizza rule}) and are composed by \textit{high-qualified engineers} and \textit{T-shape people}. These teams promote \textit{skills over roles}, \textit{leadership from management}, and more frequently \textit{single management} (referred to as product manager, product leader, technical leader, etc.) versus \textit{multiple management}. Product teams may involve those skills related to analysis, architecting and design, development, testing, operation (system administration, monitoring), security, etc. or collaborate with other teams/departments that own some of these skills. The data show that some participating organizations like ID02 differentiate \textit{first-level operations} from  \textit{second-level operations}. The first one is a 24x7 operation that mainly offers monitoring, alerting, and supporting of the software into production; the second one works closely with developers from the beginning of product development by establishing non-functional requirements (\textit{NFR shared responsibility}), configuration files, deployment scripts, and other activities related to operations.

\vspace{0.2cm}
\noindent \fbox{\begin{minipage}{8.5cm}
\textit{ID02 “Operations is divided: a 24x7 operating room called first-level of operations, and a second-level of operations that are common and cross to product teams. The first level handles systems into production while the second level handles everything from development to production. Engineers of the second-level are a very valuable resources and have a high cost, much more than the first level technicians. So when we implement new features in production we look for up to 80\% of operations to be performed by the first-level ---if they were not previously automated---, and the remaining 20\% are performed by the second-level operations.”}
\end{minipage}}
\vspace{0.2cm}

While using these codes we realized that \textit{self-organization} and \textit{autonomy} are subjective constructs that depend on the background of the interviewees. We wrote some memos to clarify how SE literature defines these constructs and compared all codings to check if this meaning fit with the narrative of the interviews. The memo is as follows:

\vspace{0.2cm}
\noindent \fbox{\begin{minipage}{8.5cm}
MEMO: \textit{Self-organizing teams are composed of “individuals [that] manage their own workload, shift work among themselves based on need and best fit, and participate in team decision making” (Highsmith 2004), %have common focus, mutual trust, respect, and the ability to organize repeatedly to meet new challenges (Cockburn and Highsmith 2001), 
and exhibit \textbf{autonomy} (Takeuchi and Nonaka 1986), i.e. decision making authority is brought to the level of operational problems (Moe and Dingsoyr 2008). %Self-organizing teams have been identified as one of the critical success factors of projects (Chow and Cao 2008). 
[Hoda, 2013]}
\end{minipage}}
\vspace{0.2cm}

Some participating organizations highlighted that product teams have \textit{external dependencies} with other teams, mainly architecture, quality assurance, system administration, database administration, security, and first-level operations dependencies. These dependencies usually generate \textit{organizational barriers} due to poor communication and lack of collaboration. Some other organizations, although they tried to face with these organizational barriers, still show \textit{cultural barriers} mainly between developers and operators (sometimes due to previous organizational silos that remain as vestigial cultural silos). Both organizational and cultural barriers are related to \textbf{silos}, which are instantiated as: \textit{operation silos}, \textit{system administration silos}, \textit{security silos}, \textit{quality assurance silos}, \textit{architecture silos}, and so on.

As we were characterizing product teams, we realized of some initial structures according to their maturity: 

{\setlength{\parindent}{0pt} \textbf{Emerging product teams} resulting from the eventual inter-departmental collaboration between dev \& ops and showing organizational silos.}

\vspace{0.2cm}
\noindent \fbox{\begin{minipage}{8.5cm}
\textit{ID01 “There is an organizational structure of departments so that development and operation still belong to different departments. We are halfway there.”}
\end{minipage}}
\vspace{0.2cm}

{\setlength{\parindent}{0pt} \ \textbf{Stable product teams} resulting from the creation of teams in which developers and operators daily collaborate, but there exist a transfer of work between them, showing some cultural barriers.}

\vspace{0.2cm}
\noindent \fbox{\begin{minipage}{8.5cm}
\textit{ID14 “High degree of collaboration although there are always silos, such as security, or others that have more to do with people, culture, the management of infrastructure, or human resources.”}
\end{minipage}}
\vspace{0.2cm}

{\setlength{\parindent}{0pt} \ \textbf{Consolidated product teams}, which have dealt both organizational and cultural silos by aligning dev \& ops goals with business goals and show cross-functional teams with shared product ownership, end-to-end product vision and high-levels of self-organization and autonomy. }

\vspace{0.2cm}
\noindent \fbox{\begin{minipage}{8.5cm}
\textit{ID23  “I would say that there are not many silos. We work as a global project, we all know our tasks and how they are related to each other.”}
\end{minipage}}
\vspace{0.2cm}

We also realized that these product teams are supported by \textbf{horizontal (cross) teams}, which may provide: 

{\setlength{\parindent}{0pt}  \textit{- Platform servicing} (e.g., CI/CD and realising tools) and infrastructure (e.g., cloud infrastructure, virtualization or containerization, etc.) to implement \textbf{best practices}, such as continuous integration, continuous testing, continuous delivery and deployment, infrastructure as code, and continuous monitoring.}

{\setlength{\parindent}{0pt} \textit{- Evangelization and mentoring} on DevOps practices for promoting \textbf{culture values}, such as \textit{communication}, \textit{transparency}, and \textit{knowledge sharing}.}
    
{\setlength{\parindent}{0pt} \textit{- Rotary human resources}, i.e.,  horizontal teams may facilitate and provide product teams with human resources when these teams lack of specific skills to undertake and accomplish their work and implement best practices.}

\vspace{0.2cm}
\noindent \fbox{\begin{minipage}{8.5cm}
\textit{ID2 (horizontal team) “We adopted a DevOps strategy implemented through a DevOps cross team of approximately 15-25 people (may vary over time) that supports 4 development areas (.es, Marketplace, Food and Tickets) in which there are almost 100 development people (approx. 20-30 people per area). The engineers of the cross team are assigned within the squads of the development areas as needed for a period of time. For example, at a given time an .es team needs two developers, then it needs two more system engineers, and so on. These engineers help squads to implement DevOps best practices. [...]. The cross team is not large enough to permanently assign these resources to the development areas.”}
\end{minipage}}
\vspace{0.2cm}

While using these codes we also realized that there were different kind of horizontal teams such as \textit{DevOps Center of Excellence (DevOps CoE)}, \textit{DevOps chapter} and \textit{Platform team}. Despite there are some differences among these teams, all of them refer to the same construct, i.e., teams that provides platform, infrastructure, IT operation, and/or mentoring among others, as a service. This gives autonomy to product teams. We wrote some memos to clarify this meaning as follows:

\vspace{0.2cm}
\noindent \fbox{\begin{minipage}{8.5cm}
MEMO: \textit{Horizontal (cross) DevOps teams, either DevOps CoE, DevOps chapter, or Platform team, aim to provide a DevOps platform,  IT operation, or mentoring, for autonomous product teams. They own DevOps skills \& culture, platform, tools, and infrastructure to provide product teams with (i) servicing of CI/CD platform and environments for dev, test, or even pre-production, (ii) mentoring and evangelizing, and sometimes  (iii) engineers who get involved in product teams with exclusive dedication but limited in time until these teams are capable of the “you build it, you run it”.}
\end{minipage}}
\vspace{0.2cm}

\begin{figure}[!h]
\centering
\includegraphics [width=7.5cm]{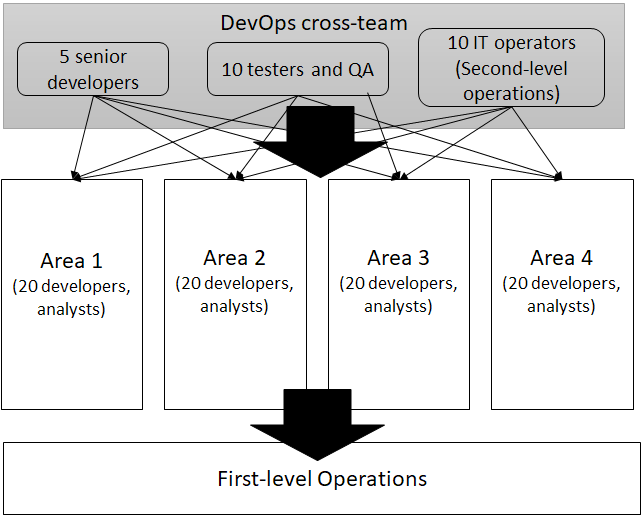}
\caption{Organizational team structure by ID2}
\label{fig:ejemplo1}
\end{figure}

\begin{figure}[!h]
\centering
\includegraphics [width=8cm]{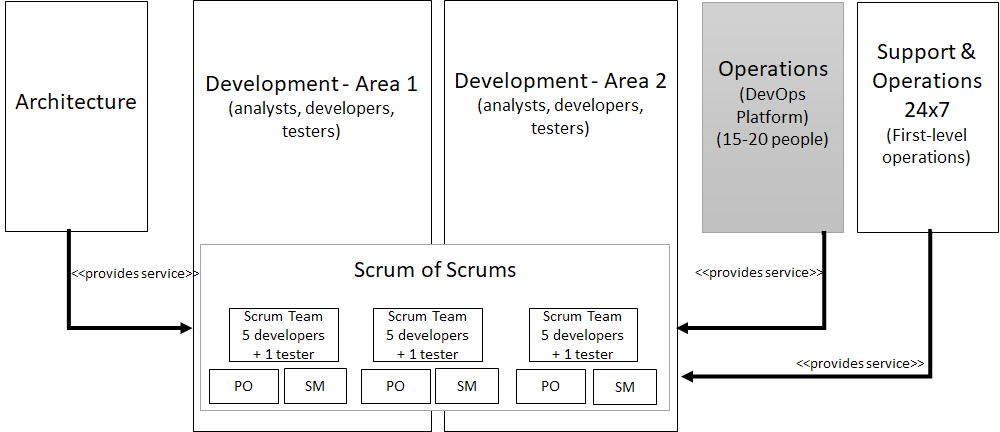}
\caption{Organizational team structure by ID11}
\label{fig:ejemplo2}
\end{figure}

\begin{figure}[!h]
\centering
\includegraphics [width=7.5cm]{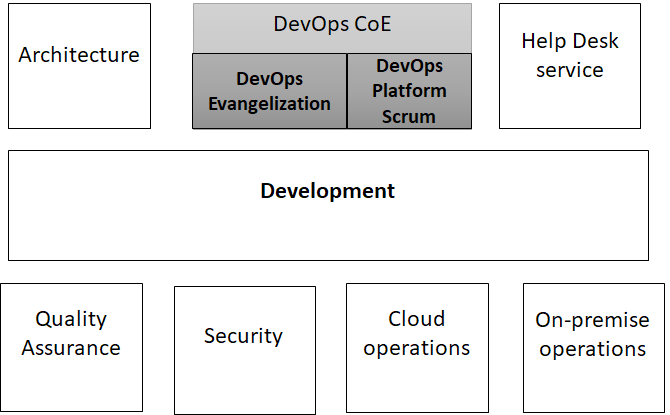}
\caption{Organizational team structure by ID17}
\label{fig:ejemplo3}
\end{figure}

Figures \ref{fig:ejemplo1}-\ref{fig:ejemplo3} shows different occurrences of these horizontal teams. Hence, Figure~\ref{fig:ejemplo1} shows a cross team composed of high qualified engineers on DevOps culture, specifically 5 senior developers, 10 testers and quality assurance engineers, and 10 IT operators, who get involved in product teams. These engineers are involved in product teams with exclusive dedication but limited in time, until product teams are capable of doing all their responsibilities, from planing, analysis, development, testing, deployment, to operation. This means that these horizontal teams are composed of engineers that move through the product teams according to their needs. The reason why these engineers are not part of the product teams is that these organizations (like ID2) do not have  human resources enough to involve the necessary engineers in all the product teams.  

Figure~\ref{fig:ejemplo2} shows an example in which the operations department assumes the DevOps culture, provides to developers (Scrum teams) with platform and infrastructure, and enables scrum teams to be autonomous. This means, the operations department assumes the functions of a DevOps platform team. This example differs from the previous one in the fact that there is no immersion of engineers from the horizontal team to the product teams.

\vspace{0.2cm}
\noindent \fbox{\begin{minipage}{8.5cm}
MEMO: \textit{When talking about platform, interviewees mainly refer to platforms that implement CI \& CD mechanisms and enable product teams to be autonomous and reach the principle “you build it, you run it”.}
\end{minipage}}
\vspace{0.2cm}

Figure~\ref{fig:ejemplo3} shows another approach in which a horizontal team (i) develops, in collaboration with the rest of departments, a DevOps platform for internal use, and (ii) evangelizes  DevOps practices. This example differs from the previous ones in the fact that the horizontal team behaves as a product team (the product is the DevOps platform) while provides service to both product teams and classical operations (either cloud or on-premise).

\vspace{0.2cm}
\noindent \fbox{\begin{minipage}{8.5cm}
[ID17] \textit{“The DevOps department is composed of two squads. The first one automates processes and develops a DevOps platform for internal use (by dev, ops, arch, qa, sec), and the second one is a chapter that works closely with other departments [development, cloud operations, on-premise operations, security, architecture], evangelizing both DevOps practices and the use of the internal platform.”}
\end{minipage}}
\vspace{0.2cm}

Finally, the following excerpts show more evidence of the existence of horizontal DevOps teams in the participating organizations and its importance to make product teams autonomous. 

\vspace{0.2cm}
\noindent \fbox{\begin{minipage}{8.5cm}
[ID23] \textit{“a highly qualified senior team prepares and provides a DevOps platform that is consumed by the development teams.”}
\end{minipage}}
\vspace{0.2cm}

\vspace{0.2cm}
\noindent \fbox{\begin{minipage}{8.5cm}
[ID28] \textit{“A cross team provides service to five product teams so that they are able to deploy on their own, although if they occasionally need support, the cross team gets involved. For example, for certain very specific cases in which more support or expertise in databases, elastic search, or any other infrastructure service was needed, my team [the cross team] got involved.”}
\end{minipage}}
\vspace{0.2cm}

\vspace{0.2cm}
\noindent \fbox{\begin{minipage}{8.5cm}
[ID29] \textit{“ID29 has evolved into a value streams model, where products are organized around business Product Owners. Platform/enabling teams were created organized around technical capabilities that collaborate with the different value streams with the goal of making them independent and maturing their DevOps capabilities.”}
\end{minipage}}
\vspace{0.2cm}

%%%%%%%%%%%%%%%%%%%%%%%%%%%%%%%%%%%%%%%%%%%%%%%%%%%%%%%%%%%%%%
\subsection{Category selection: key variables}

%Once the initial coding phase is done and the core categories have been established, we proceed with the \textbf{focused coding} phase as follows. An exhaustive review of the initial coding allowed us to identify the most relevant codes for the topic at hand. One more time, we have highlighted the codes from the initial coding along with raw data quotes to make explicit the chain of evidence.

We found in the raw data that certain categories were especially important when starting to sketch organizational team structures, which we refer to as variables. These variables are as follows: 

\subsubsection{Product ownership sharing}
We observed that there exist a relationship between how product teams share the product ownership and how these teams are structured. For example, ID29 shows  a high-level of sharing of the product ownership within product teams, which are cohesive, small (less than 12 people) and multidisciplinary. On the contrary,  the interviewee of ID02 admits that there is room for improvement due to the strong separation of tasks within the teams and the existence of post-mortem meetings to seeking blames. In this way, we were able to establish three levels on the \textbf{shared ownership variable}: crawl, walk and run.  The following excerpts show some evidence:

\vspace{0.2cm}
\noindent \fbox{\begin{minipage}{8.5cm}
[ID29] \textit{“We were responsible for the ownership of all the digital asset, so much that the platform was down on July 4Th and many people from different "areas" left their barbecues [they were in United States celebrating the national holiday] to fix the problem. I detected the issue and immediately a Slack call with 11 people started. Everyone was clear that it was not a Front problem, a Back problem or an Ops problem, but it was a problem of the team and everyone was committed to solving it.”}
\end{minipage}}
\vspace{0.2cm}

\vspace{0.2cm}
\noindent \fbox{\begin{minipage}{8.5cm}
[ID02] \textit{“We increasingly share the responsibility of delivering new features and functionality, but we still have a long way to go, because we come from a culture that is so siloed that it is very difficult for us to share that responsibility.”}
\end{minipage}}
\vspace{0.2cm}

\subsubsection{Leadership from management}
We realized that shared ownership is highly related to leadership from management, which is an interesting variable to be examined due to its impact on organizational team structures. Hence, non-shared leadership usually leads to non-shared ownership because, if there are multiple managers within the same team (typically one development manager and one operation manager), then it is difficult for all members to feel the product as a whole. This means, each member tends to take only a part of the responsibility (developers give priority to develop new features whereas operators give priority to service stability). Organizations like ID16 face the problems arising from this situation, while many others like ID08 and ID21 take for granted the need for product teams to have only one leader from management, even if there are major development and operations areas at organizational level. In this way, we established two levels on the \textbf{leadership from management variable}: single and multiple.  The following excerpts show some evidence:

\vspace{0.2cm}
\noindent \fbox{\begin{minipage}{8.5cm}
[ID02] \textit{“In most cases the development and operations staff involved in a product have different managers. However, in new developments related to mobile applications we are trying to avoid this, reduce dependencies and improve collaboration.”}
\end{minipage}}
\vspace{0.2cm}

\vspace{0.2cm}
\noindent \fbox{\begin{minipage}{8.5cm}
[ID08] \textit{“Although structurally within the company the development and operations staff depends on different areas, in each product team there is a single leader whose responsibility is to bring that team together.”}
\end{minipage}}
\vspace{0.2cm}

\vspace{0.2cm}
\noindent \fbox{\begin{minipage}{8.5cm}
[ID21] \textit{“There is no such distinction between development and operations staff. In our team we all have the same leader and only report to him/her.”}
\end{minipage}}
\vspace{0.2cm}

\subsubsection{Organizational silos \& Cultural silos}
We also found that the existence of strong hierarchical organization charts and departmental structure impact the structure of teams because both organizational and cultural silos undermined the adoption of DevOps practices and culture. The structure of some organizations like ID01 leads to the creation of silos and find themselves with serious problems to adopt DevOps. In other cases like ID27 the organizational silos were broken, but the cultural ones remain (at least for a while) hindering the DevOps adoption. Many organizations like ID29 have managed to transform their structure, eliminated all the silos and achieved a complete adoption of DevOps. Recently founded organizations like ID09 usually do not face silo problems because they were born with a structure that favors DevOps. By observing the big picture, we established two levels on the \textbf{organizational silo variable}: yes, no; and three levels on the \textbf{cultural silo variable}: yes, no, vestigial (previous silos remain as vestigial cultural silos). The following excerpts show some evidence:

\vspace{0.2cm}
\noindent \fbox{\begin{minipage}{8.5cm}
[ID01] \textit{“Today there is an organizational structure of departments so that development and operation still belong to different departments. There is a wall between both areas that prevent DevOps adoption. So, until now, there are organizational silos, and therefore, cultural silos too.”}
\end{minipage}}
\vspace{0.2cm}

\vspace{0.2cm}
\noindent \fbox{\begin{minipage}{8.5cm}
[ID27] \textit{“I was sure that the adoption would only be viable at the moment we managed to either break the silos or float above them. So we created mixed and autonomous teams with people from different departments. However, we still talk about the product team, the operation team, the architecture team, the QA team,...”}
\end{minipage}}
\vspace{0.2cm}

\vspace{0.2cm}
\noindent \fbox{\begin{minipage}{8.5cm}
[ID29] \textit{“Currently the whole IT department works according to the DevOps culture. The continuous growth has led ID29 to a deep organizational transformation.”}
\end{minipage}}
\vspace{0.2cm}

\vspace{0.2cm}
\noindent \fbox{\begin{minipage}{8.5cm}
[ID09] \textit{“We only have product teams with end-to-end responsibility of an integrated product. Since we were formed in 2010, we did not have the traditional culture of IT departments so it was not so complicated to apply DevOps.”}
\end{minipage}}
\vspace{0.2cm}

\subsubsection{Collaboration}
We also noticed that collaboration frequency is highly related to the team structures and is a critical variable for DevOps adoption. Indeed collaboration is one of the key values of DevOps culture. Hence, the members of product teams may work together regularly on a daily basis to undertake all the product life-cycle, as it happens in organizations like ID23. This implies a daily collaboration between team members and usually a daily meeting, without detriment of other less frequent meetings with other related teams. However, the members of product teams in other organizations have more differentiated roles (dev versus ops) so that they work together but in different tasks. This means, there is not a real collaboration, but a transfer of responsibilities, as it happens in organizations like ID05. In these cases the collaboration frequency is on a weekly basis or even more. In this way we established three levels on the \textbf{collaboration frequency variable}: daily, frequent and eventual.  The following excerpts show some evidence:

\vspace{0.2cm}
\noindent \fbox{\begin{minipage}{8.5cm}
[ID23] \textit{“There are daily meetings within each product team and everyone collaborates with their colleagues in their day to day. In addition there are other meetings typically performed in Scrum or SAFE with other teams to have absolute transparency with the entire project.”}
\end{minipage}}
\vspace{0.2cm}

\vspace{0.2cm}
\noindent \fbox{\begin{minipage}{8.5cm}
[ID05] \textit{“Collaboration between development and operations is done through a ticketing application, although we share physical work space and sometimes work together.”}
\end{minipage}}
\vspace{0.2cm}

\subsubsection{Autonomy}
Last but not least, we found that autonomy might reveal the organizational team structure of a company. We understand that a DevOps product team is entirely autonomous when it does not have external dependencies to fulfill its responsibilities, this implies having an end-to-end vision and taking complete charge of a product from its conception and implementation to its deployment and monitoring. This is hard to achieve and we only found a few organizations like ID03 whose product teams implement the practice \textit{continuous deployment} and \textit{continuous feedback}, and thus, being completely autonomous. The most common practice is that  product teams implement \textit{continuous delivery} so that they can deploy in a pre-production environment, but they need external approval to go into production. These approvals may come directly from business or from technical areas such as quality or security. This was very usual in most organizations like ID08, even if their DevOps maturity was high. However, in some organizations, product teams still have many dependencies and they do not manage continuous delivery, much less continuous deployment. For example, we found organizations like ID01 where product teams do not yet have the ability to create their own environments and still depend on an operations department. In the case of these organizations there is still a long way to go in adopting DevOps. In this way, we established three levels on the \textbf{autonomy variable}: high (no dependencies), medium (deployment dependencies) and low (many dependencies, i.e. dependencies beyond deployment).  The following excerpts show some evidence:

\vspace{0.2cm}
\noindent \fbox{\begin{minipage}{8.5cm}
[ID03] \textit{“There are no external dependencies to deploy in production. If the Product Owner and the team as a whole consider that a feature is correct, then it is deployed in production.”}
\end{minipage}}
\vspace{0.2cm}

\vspace{0.2cm}
\noindent \fbox{\begin{minipage}{8.5cm}
[ID08] \textit{“Yes, of course, there are dependencies to make deployments in production as it requires the approval of the service manager and the customer. However, we work free of dependencies with a pre-production environment very similar to a production environment and when authorized we quickly deploy in production.”}
\end{minipage}}
\vspace{0.2cm}

\vspace{0.2cm}
\noindent \fbox{\begin{minipage}{8.5cm}
[ID01] \textit{“Before, Dev did not even have visibility of the deployment process and when an environment for a new application was requested, the task was done behind the wall. When Ops created or configured the environment, it might not fit with what Dev needed and a lot of time was lost in this process. As now the collaboration is higher and people from Dev and Ops meet every 2 days, the requests of environments are more agile and effective.”}
\end{minipage}}
\vspace{0.2cm}

Finally, Table~\ref{tab:variables} summarizes the six key variables that allow us to characterize and categorize team structures and lay the foundation to build a cohesive theory. 

\begin{table}[!h]
    \centering
    \caption{Variables used for the focused coding}
    \begin{tabular}{p {2cm} p {3.8cm} p {1.6cm}}
         \hline 
         \textbf{Variable} & \textbf{Meaning} & \textbf{Value range}\\\hline
         Leadership from management & Both Dev and Ops are under the same management in a product team & Single \newline Multiple \\\hline
         Product \newline ownership \newline sharing & Dev and Ops share responsibility for delivering value & Crawl\newline Walk\newline Run\\ \hline
         Collaboration frequency & Dev and Ops work and meet together regularly & Eventually \newline Frequent \newline    Daily\\\hline
         Organizational silos & The structure of the organization causes silos & Yes \newline No \\ \hline
         Cultural silos & Prior work habits and processes causes silos between people & Yes \newline No \newline  Vestigial\\\hline
         Autonomy & Product teams have no external dependencies to perform all its responsibilities & Low  \newline Medium  \newline High  \\\hline 
    \end{tabular}
    \label{tab:variables}
\end{table}

%%%%%%%%%%%%%%%%%%%%%%%%%%%%%%%%%%%%%%%%%%%%%%%%%%%%%%%%%%%%%%%%%%%%%%%%%%%%%
\subsection{Building a Theory}

We followed the guidelines for describing a theory stated by Sj{\o}berg et al. \cite{sjoberg:2008}, according to which a theory description should be divided into four parts: \textit{constructs} (what are the basic elements) i.e., categories and variables; \textit{propositions} (how do the constructs interact), i.e., relationships between constructs and variables; \textit{explanations} (why are the propositions as specified); and \textit{scope} (what is the universe of discourse in which the theory is applicable, which was already described in Section~\ref{sec:methodology}). Hence, we built relationships between the above mentioned categories and variables, and integrate them into a conceptual framework. The memos previously taken were essential in this process, in which it was necessary to keep in mind all the information from the analyzed interviews. Thus, we started to build a first approach of the conceptual framework as follows.

\begin{itemize}
    \item []\textbf{Constructs}
    \item [C1.] Product teams: emerging teams, stable teams, and consolidated teams. 
    \item [C2.] Horizontal DevOps teams: centers of excellence, chapters, and platform teams. 
    \item [C3.] Best practices: continuous everything (testing, integration, delivery, deployment, monitoring and feedback) and infrastructure as code. 
    \item [C4.] Cultural values: sharing, transparency, and communication.  
    \item [C5.] Key variables: leadership, product ownership, collaboration, silos, and autonomy.
\end{itemize}

\begin{itemize}
    \item []\textbf{Propositions}
\item [P1.1] Product teams are often composed of high qualified and T-shape engineers in small teams, and are often characterized by an end-to-end product vision, multidisciplinarity, self-organization. 

\item [P1.2] Product teams may be characterized, according to their maturity (emerging, stable, and consolidated), with different levels of shared product ownership and autonomy. 

\item [P1.3] Product teams put skills ahead of roles and, according to their maturity (emerging, stable, and consolidated teams) prioritize single leadership over multiple management.

\item [P2.1] Horizontal DevOps teams are instantiated in companies through centers of excellence, chapters, and platform teams. CoE and platform teams are external teams, whereas DevOps chapters designate people having DevOps skills and working within the same general competency area across different product teams.

\item [P2.2] Horizontal DevOps teams may provide product teams with platform servicing to support them with the appropriate technologies and tools to automate the release to production.

\item [P2.3] CoE evangelize and mentor product teams in a continuous improvement process and DevOps maturity through cultural values and best practices. 

\item [P2.4] CoE may provide rotary human resources to product teams to boost teams with the appropriate skills to release to production.

\item [P3.] Product teams can implement the best practices due to the support of horizontal DevOps teams.

\item [P4.] Product teams can introduce, settle and/or enhance the cultural values due to the support of horizontal DevOps teams.

\item [P5.] Different configurations of the key variable values result in different DevOps maturity levels: emerging, stable, and consolidated.
%P5. A first-level 24x7 operations, which mainly are in charge of monitoring, alerting, and supporting software into production, coexist with product teams.
\end{itemize}

\begin{itemize}
 \item []\textbf{Explanations} 
\item [E1.1] Emerging product teams are characterized by eventual interdepartmental collaboration between dev \& ops which is hampered by organizational silos. 

\item [E1.2] Stable product teams are characterized by dev \& ops daily work but there exist transfer of work between them due to cultural barriers that persist. 

\item [E1.3] Consolidated product teams are characterized by having a shared product ownership, an end-to-end product vision, multidisciplinarity, and all the skills to do their work resulting in high levels of self-organization and autonomy.

%E5.1. Emerging product teams seem to have a low performance.

%E5.2. Stable product teams seem to have a medium performance.

%E5.1. Consolidated product teams seem to have a high performance.
\end{itemize}

% \textbf{Scope} 
%Organizations that have been adopting DevOps for at least two years.

\begin{figure*}[!h]
\centering
\includegraphics [width=18cm]{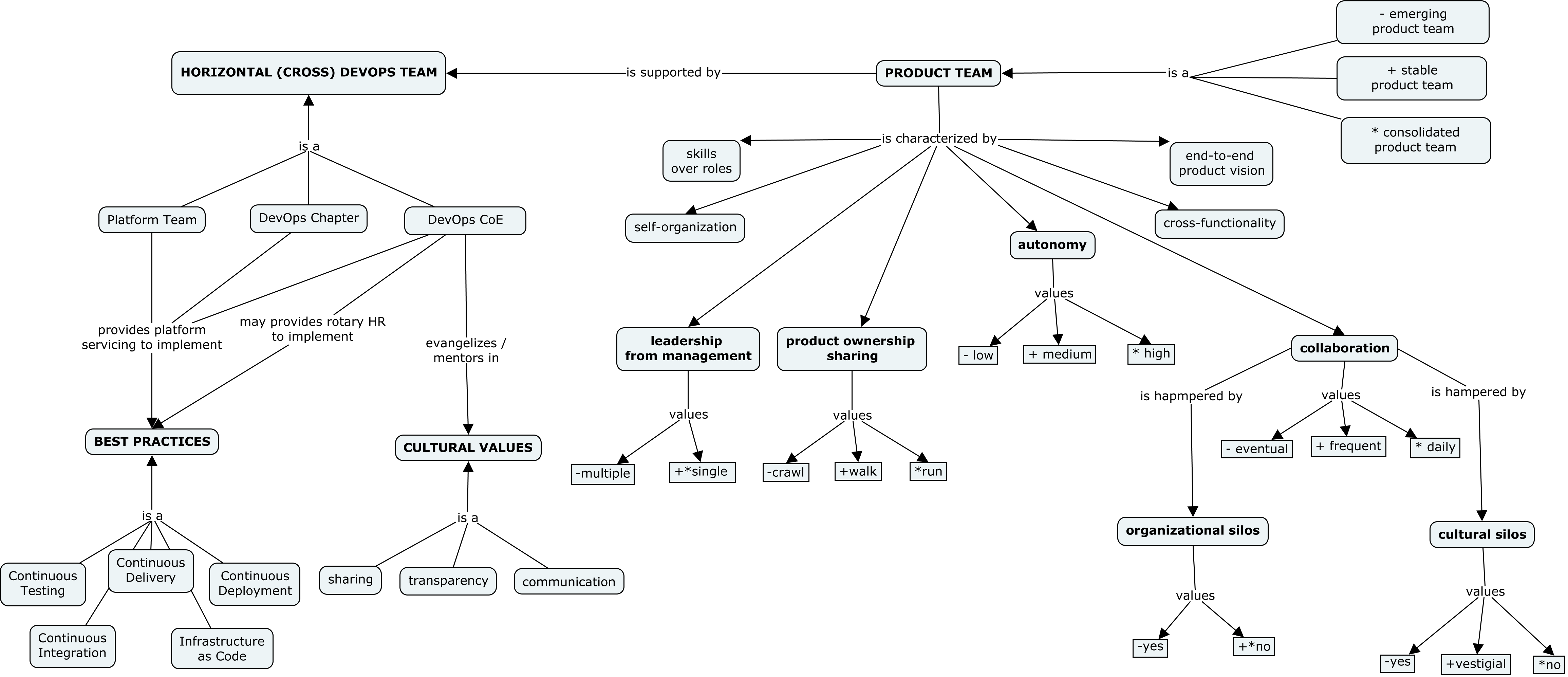}
\caption{Conceptual framework for DevOps organizational team structures}
\label{fig:model}
\end{figure*}

Figure~\ref{fig:model} graphically shows the conceptual framework. Notice we used ovals for categories and variables, and rectangles for values of variables. Notice that we used the symbols - + and * to denote the values for emerging, stable and consolidated product teams respectively. 

%%%%%%%%%%%%%%%%%%%%%%%%%%%%%%%%%%%%%%%%%%%%%%%%%%%%%%%%%%%%%%%%%%%%%%%%%%%%%%%%%%%%%%%%%%%%%%%%%%%%
%%%%%%%%%%%%%%%%%%%%%%%%%%%%%%%%%%%%%%%%%%%%%%%%%%%%%%%%%%%%%%%%%%%%%%%%%%%%%%%%%%%%%%%%%%%%%%%%%%%%
\section{A taxonomy of DevOps Team Structure Patterns (RQ1)}
\label{sec:model}

The conceptual framework resulting from the previous section provides the basis to create a taxonomy of DevOps team structure patterns. Section 4.1 describes these patterns focusing on the resulting product teams; Section 4.2 describes the role of horizontal cross teams; finally Section 4.3 depicts a global vision of this taxonomy. 

\subsection{Patterns description}

The taxonomy of team structure patterns is the result of (i) analyzing the six variables of the conceptual framework (see Table~\ref{tab:variables}), and (ii) assigning a value for each organization. This means, from qualitative data (the interview transcriptions) we made an effort of interpretation and quantified these variables. Table~\ref{tab:organizationscharacteristics} shows this relation, i.e., organization ID, variable, and value. Later, the analysis of the common factor of the variables values allowed us to identify four patterns of team structures (see Table \ref{tab:patterns}), which show similar characteristics regarding how they work, communication pathways, and team performance (i.e., collaboration, autonomy, product ownership sharing, leadership, and organizational or cultural silos). These patterns are described as follows.

\begin{table*}[!h]
    \centering
    \caption{Organizations characteristics analysis}
    \begin{tabular}{p{1cm} p{2cm} p{2cm} p{2cm} p{2cm} p{2cm} p{3cm}}
         \hline \textbf{ID} & \textbf{Leadership from management} & \textbf{Shared ownership} & \textbf{Collaboration frequency} & \textbf{Organizational silos} & \textbf{Cultural silos} & \textbf{Autonomy} \\\hline
         ID1 & Multiple & Crawl & Frequent & Yes & Yes & Low \\
         ID2 & Single & Walk & Daily & No & Vestigial & Medium \\
         ID3 & Single & Run & Daily & No & No & High \\
         ID4 & Multiple & Walk & Frequent & Yes & Yes & Low \\
         ID5 & Multiple & Crawl & Eventual & Yes & Yes & Low \\
         ID6 & Single & Crawl & Frequent & No & Yes & Low \\
         ID7 & Single & Crawl & Frequent & No & Yes & Medium\\
         ID8 & Single & Run & Daily & No & Yes & Medium\\
         ID9 & Single & Run & Daily & No & No & High\\
         ID10 & Single & Run & Daily & No & No & Medium\\
         ID11 & Single & Run & Daily & No & Vestigial & Medium\\
         ID12 & Multiple & Crawl & Eventual & Yes & Yes & Low\\
         ID13 & Single & Run & Daily & No & No & High\\
         ID14 & Single & Walk & Frequent & No & Yes & Low\\
         ID15 & Single & Run & Daily & No & No & Medium\\
         ID16 & Multiple & Crawl & Frequent & Yes & Yes & Low\\
         ID17 & Single & Run & Daily & No & Vestigial & Medium\\
         ID18 & Single & Run & Daily & No & Yes & Medium\\
         ID19 & Single & Walk & Frequent & No & Yes & Medium\\
         ID20 & Multiple & Crawl & Eventual & Yes & Yes & Low\\
         ID21 & Single & Walk/Run & Daily & No & No & Medium\\
         ID22 & Single & Run & Daily & No & No & Medium\\
         ID23 & Single & Run & Daily & No & No & Medium\\
         ID24 & Single & Walk/Run & Daily & No & Vestigial & Medium\\
         ID25 & Multiple & Crawl & Frequent & Yes & Yes & Low\\
         ID26 & Single & Walk/Run & Daily & No & No & Medium\\
         ID27 & Single & Walk & Frequent & No & Yes & Medium\\
         ID28 & Single & Run & Daily & No & No & Medium\\
         ID29 & Single & Run & Daily & No & No & Medium\\
         ID30 & Single & Run & Daily & No & No & Medium\\
         ID31 & Single & Run & Daily & No & No & Medium\\\hline
    \end{tabular}
    \label{tab:organizationscharacteristics}
\end{table*}

\begin{table*}[!h]
    \centering
    \caption{Taxonomy of DevOps Team Structure Patterns: Organizations' classification}
    \begin{tabular}{p{2cm} p{2.5cm} p{1.6cm} p{1.6cm} p{1.6cm} p{1.6cm} p{1.6cm} p{1.6cm} p{1.8cm}}
    \hline \textbf{Name} & \textbf{Organizations} & \textbf{Leadership from management} & \textbf{Shared ownership} & \textbf{Collaboration frequency} & \textbf{Organizational silos} & \textbf{Cultural silos} & \textbf{Autonomy} \\\hline
    Interdepartmental Dev \& Ops collaboration \newline & ID1, ID4, ID5, ID12, ID16, ID20, ID25 & Multiple & Crawl & Eventual/ Frequent & Yes & Yes & Low \\
    Interdepartmental Dev-Ops team \newline & ID6, ID7, ID14, ID19, ID27 & Single/ Multiple & Crawl/Walk & Frequent & No & Yes & Low/Medium \\
    Boosted (cross-functional) DevOps team \newline & ID2, ID11, ID13, ID17, ID24, ID30 & Single & Walk/Run & Daily & No & No/Vestigial & Medium/High \\
    Full (cross-functional) DevOps team & ID3, ID8, ID9, ID10, ID15, ID18, ID21, ID22, ID23, ID26, ID28, ID29, ID31 & Single & Run & Daily & No & No & Medium/High \\\hline
    \end{tabular}
    \label{tab:patterns}
\end{table*}

\subsubsection{Pattern A: Interdepartmental Dev \& Ops collaboration}
%The first team pattern is named \textbf{interdepartmental Dev \& Ops collaboration}. In this case 
According to this pattern, people from dev and ops who belong to different departments collaborate sporadically in a specific project, but they do not form a stable team. As dev and ops belong to different departments and the collaboration is not stable, they have different managers, the product ownership is not actually shared, and the silos are a major barrier. Developers and operators only collaborate occasionally (on a monthly or at most weekly basis), each one works on their tasks. This means, there exist a transfer of tasks, no collaboration. Furthermore, the team has no autonomy as responsibilities are delegated and there are dependencies with other teams. In short, the characteristics of this organizational team pattern are as follows:

\begin{itemize}
    \item [--] Leadership from management: Multiple
    \item [--] Shared product ownership: Crawl
    \item [--] Collaboration frequency: Eventual to Frequent
    \item [--] Organizational silos: Yes
    \item [--] Cultural silos: Yes
    \item [--] Autonomy: Low
\end{itemize}

\subsubsection{Pattern B: Interdepartmental Dev-Ops team}
%The second pattern is named \textbf{interdepartmental Dev-Ops team}. In this case 
According to this pattern, people from dev and ops form a stable product team, but they may still belong to different departments. Consequently, they may have one product manager, but often a department manager with different goals. Dev \& ops tasks are clearly separated, and although organization silos have been addressed, cultural silos still remain. However, they frequently collaborate beyond a transfer of tasks, start to share the product ownership, and begin to gain autonomy as a team. The team begins to adopt DevOps practices and principles, although in a very incipient phase. In summary, the characteristics of this organizational team pattern are as follows:

\begin{itemize}
    \item [--] Leadership from management: Single (product manager) but multiple (department managers)
    \item [--] Shared product ownership: Crawl to Walk
    \item [--] Collaboration frequency: Frequent
    \item [--] Organizational silos: No
    \item [--] Cultural silos: Yes
    \item [--] Autonomy: Low to Medium
\end{itemize}

\subsubsection{Pattern C: Boosted cross-functional DevOps team}
%The third pattern is named \textbf{boosted cross-functional DevOps team}. In this case 
According to this pattern, traditional dev teams are boosted by DevOps experts who are highly involved in these teams until they reach the "you build it, you run it" capability, thus becoming full DevOps teams. This kind of teams have a single manager and share the ownership of the product/service they are delivering. They work side by side on a daily basis and they have a medium-level of autonomy or even higher. Normally there are no cultural silos, although on some occasions there are vestigial silos derived from those that have existed previously for many years. If these teams are willing to learn from the DevOps experts who get involved with them, they will achieve the "you build it, you run it" capability thus becoming a full DevOps team. In short, the characteristics of this organizational team pattern are as follows:

\begin{itemize}
    \item [--] Leadership from management: Single
    \item [--] Shared product ownership: Walk to Run
    \item [--] Collaboration frequency: Daily
    \item [--] Organizational silos: No
    \item [--] Cultural silos: No or Vestigial
    \item [--] Autonomy: Medium to High
\end{itemize}

\subsubsection{Pattern D: Full cross-functional DevOps team}
%The fourth and last pattern is named \textbf{full cross-functional DevOps team}. In this case
According to this pattern, product teams are poly-skilled, being able to face an end-to-end product development, i.e., "you build it, you run it". These teams have a single manager and share the ownership of the product/service they are delivering. They also work side by side on a daily basis with a medium-to-high level of autonomy. It is complicated to achieve total autonomy in certain types of businesses (e.g., banking, telecommunications, retail, etc.), even for this kind of poly-skilled teams. The reason behind is that, sometimes, the deployment requires external validations such as those related to security. However, some full DevOps teams have security abilities becoming DevSecOps teams. In summary, the characteristics of this organizational team pattern are as follows:

\begin{itemize}
    \item [--] Leadership from management: Single
    \item [--] Shared product ownership: Run
    \item [--] Collaboration frequency: Daily
    \item [--] Organizational silos: No
    \item [--] Cultural silos: No
    \item [--] Autonomy: Medium to High
\end{itemize}

\subsection{The role of horizontal teams}

\begin{figure*}[!h]
\centering
\includegraphics [width=14.5cm]{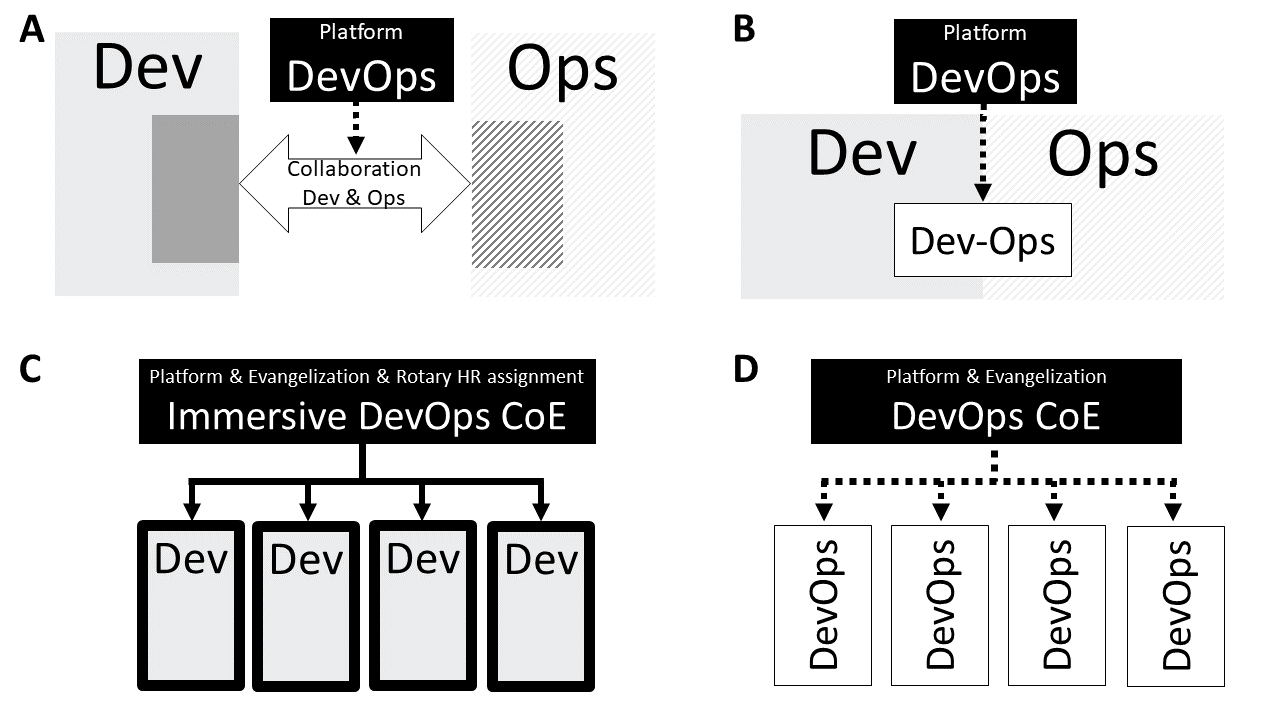}
\caption{Patterns: a) Interdepartmental Dev \& Ops collaboration, b) Interdepartmental Dev-Ops team, c) Boosted (cross-functional) DevOps team, d) Full (cross-functional) DevOps team}
\label{fig:patterns}
\end{figure*}

 Something common to all these product teams represented in the patterns is the existence of a \textbf{horizontal (cross) DevOps team}. Nevertheless, the forms in which these horizontal teams support product teams vary. How several horizontal teams come into play is described as follows. 

Pattern A and Pattern B are usually supported by \textbf{platform teams}. These horizontal teams are typically formed by people from Ops who are experts in DevOps technologies, who provide product teams the required CI/CD and monitoring platforms and infrastructure to automate DevOps practices. It can also occur that the platform is provided by a \textbf{DevOps chapter}, who is made up by DevOps experts of each product team. In both cases, the support to product teams is often limited to providing a platform. 

On the other hand, Pattern~C are supported by DevOps experts from an \textbf{immersive DevOps Center of Excellence}.  These experts provide product teams with the required platforms, as DevOps platform teams do, but they go beyond that. Moreover, the immersive DevOps CoE staff are highly involved (i.e., rotary human resource assignment) with product teams until they get the "you build it, you run it" capability and become a full DevOps team. Finally, Pattern~D are usually supported by a \textbf{DevOps Center of Excellence}, which is devoted, in addition to the provision of the DevOps platform, to the evangelization \& mentoring of the entire company in the DevOps culture and practices. 

\subsection{The four patterns in a nutshell}
Figure~\ref{fig:patterns} graphically shows the four patterns. We could say that patterns A and B are less mature than patterns C and D as product teams are not autonomous, do not share the  product ownership and business objectives, and show silos. On the contrary, patterns C and D are more mature than the previous ones as product teams are autonomous, share the product ownership and business objectives, they broke silos and are able to manage the entire product-life cycle.

\section{Statistical Analysis}
\label{sec:statistical}

This section analyzes the implications of team patterns on three well-known software delivery performance indicators (RQ2) and assesses these patterns using a quantitative analysis, which may confirm the patterns taxonomy that we previously built through qualitative analysis based on the six key variables. 

\subsection{Implications of patterns on software delivery performance (RQ2)}
\label{sec:implications}

%From this taxonomy, new research questions arise. One of these questions is about the software delivery performance that each team pattern could leverage, i.e., \textit{could different patterns leverage different software delivery performance indicators?}
This question was previously addressed by Pais \& Skelton \cite{skelton:2013}, who talk about a model for team-based software delivery performance, i.e., they agree on the fact that organizations are constrained to deliver software that reflects their team patterns and communication paths. The DevOps Research \& Assessment association (DORA) report \cite{dora:2018} also identified a set of software delivery performance profiles (elite, high, medium and low performance) and related DevOps practices to these profiles.

As the DORA report did, we gathered three software delivery performance indicators during the interviews: Lead Time (LT), i.e., the time from a change in the code is successfully running in production; Mean Time To Recovery (MTTR), i.e., elapsed time to restore a service when an incident causes its unavailability; and Deployment Frequency (DF), i.e., the number of deploys to production of an application per unit of time. Table \ref{tab:performanceresults} presents the software delivery performance indicators for each organization together with the pattern assigned to each organization according to the taxonomy described in Section~\ref{sec:model}.
 
\begin{table}[]
    \centering
    \caption{Pattern \& software delivery performance indicators}
    \begin{tabular}{p{0.7cm} p{0.8cm} p{1.3cm} p{1.3cm} p{1.8cm}}
         \hline \textbf{ID} & \textbf{Pattern} & \textbf{LT \hspace{0.5pt} (min)} & \textbf{MTTR (min)} & \textbf{DF (deploys per day)} \\\hline
         ID01 & A & 45 & 60 & 1 \\
         ID02 & C & 480 & 60 & 0.1 \\
         ID03 & D & 30 & 60 & 7.5 \\
         ID04 & A & 480 & 120 & 0.1 \\
         ID05 & A & 480 & 120 & 0.1 \\
         ID06 & B & 480 & 60 & 0.15 \\
         ID07 & B & 60 & 240 & 0.05 \\
         ID08 & D & 30 & 30 & 4.5 \\
         ID09 & D & 30 & 45 & 0.5 \\
         ID10 & D & 180 & 120 & 0.2 \\
         ID11 & C & 30 & 45 & 0.2 \\
         ID12 & A & 120 & 60 & 1 \\
         ID13 & C & 30 & 90 & 5 \\
         ID14 & B & 240 & 240 & 1 \\
         ID15 & D & 120 & 30 & 0.2 \\
         ID16 & A & 720 & 240 & 0.1 \\
         ID17 & C & 240 & 240 & 0.2 \\
         ID18 & D & 240 & 240 & 0.05 \\
         ID19 & B & 720 & 240 & 0.05 \\
         ID20 & A & 240 & 60 & 0.1 \\
         ID21 & D & 960 & 960 & 0.05 \\
         ID22 & D & 30 & 20 & 0.1 \\
         ID23 & D & 240 & 240 & 0.2 \\
         ID24 & C & 120 & 60 & 2 \\
         ID25 & A & 60 & 30 & 0.2 \\
         ID26 & D & 60 & 60 & 0.2 \\
         ID27 & B & 60 & 240 & 1 \\
         ID28 & D & 25 & 20 & 5 \\
         ID29 & D & 10 & 10 & 6 \\
         ID30 & C & 10 & 45 & 4 \\
         ID31 & D & 10 & 30 & 4 \\\hline
    \end{tabular}
    \label{tab:performanceresults}
\end{table}

From these data, we performed an in-depth statistical analysis to examine the implications of each of the patterns. First, we present the sample statistical descriptions of each indicator for each of the four patterns. Second, we study if we can assert a correlation between the pattern and the three delivery performance indicators (LT, MTTR and DF). Finally, we compare the mean of the three indicators for each of the four patterns to test significant  differences between patterns.

\subsubsection{Sample statistical description}

Table~\ref{tab:statdescription} shows the statistical descriptions of the three performance indicators for each of the four patterns. To confirm whether the sample matched the normal population, the Shapiro-Wilk tests were carried out. Based on these results, the Wilcoxon non-parametric test was used to compare the means of different samples, and the Spearman Rho test to analyze correlations. Notice that organization ID21 is not included in the statistical analysis, due to the ambiguities of some of the answers in the interview and numerical data that were clearly atypical. Finally, we have 30 samples.

\renewcommand\cellalign{rc}

\begin{table}[htbp]
  \centering
  \caption{Statistical description of delivery performance indicators per pattern}
    \begin{tabular}{|r|ccrr|}
%    \toprule
    \hline
    Indicator & Pattern & $N$ & Mean & Std. Dev. \\
    \hline
%    \midrule
    \multirow{4}{*}{\makecell{ LT \\ (Lead Time) } } & A & 7 & 306.43 & 258.049 \\
     & B & 5 & 312 & 285.867 \\
     & C & 6 & 151.67 & 182.364 \\
          & D & 12 & 83.75 & 88.347 \\
    \hline
    \multirow{4}{*}{\makecell{ MTTR \\ (Mean Time to Recovery) } } & A & 7 & 98.57 & 70.812 \\
           & B & 5 & 204 & 80.498 \\
    & C & 6 & 90 & 75.299 \\
          & D & 12 & 75.42 & 82.116 \\
    \hline
     \multirow{4}{*}{\makecell{ DF \\ (Deployment Frequency) } }  & A & 7 & 0.371 & 0.4309 \\
           & B & 5 & 0.45  & 0.5037 \\
     & C & 6 & 1.917 & 2.1470 \\
          & D & 12 & 2.371 & 2.8038 \\
    \hline
    \end{tabular}%
  \label{tab:statdescription}%
\end{table}%

\subsubsection{Correlation between patterns and performance}

We ran an exploratory analysis to detect the main correlations between the pattern (as variable) and the three software delivery performance indicators. Table~\ref{tab:Spearman} displays the results of the Rho Spearman coefficient. Notice that the pattern has a negative correlation with the indicator LT ($\textrm{Coef.}=-0.505$; $\textrm{p-value}=0.004$), which means that the more mature a pattern is, the lower (i.e., better) lead time is achieved. This correlation is highly significant, even at level $0.01$.
%This coincides with the deductions made during the qualitative analysis in which it was perceived that companies that followed more advanced organizational patterns had a higher team performance. 
Milder significant negative correlation was also obtained with the indicator MTTR ($\textrm{Coef.}= -0.382$; $\textrm{p-value}=0.037$), fulfilled at level $0.05$. However, results do not show a significant correlation between the pattern and the indicator DF ($\textrm{p-value}= 0.058$). Although the most mature patterns tend to have higher deployment frequency, it is also true that on some occasions deployment windows are fixed regardless of the adopted pattern (typically, one per sprint), so the difference observed in this indicator is not so noticeable as in the others.
\renewcommand{\arraystretch}{1.2}

\begin{table}[htbp]
  \centering
  \caption{Spearman correlation for Pattern and performance indicators}

    \begin{tabular}{lrrr}
    \hline
          & LT & MTTR & DF \\
    \hline
    \hline
    Coef. & {-0.505} & {-0.382} & 0.35 \\
    %Sig.  & 0.004 & 0.037 & 0.058 \\
    p-value  & 0.004 & 0.037 & 0.058 \\
    $N$     & 30    & 30    & 30 \\
    
    \hline
    \end{tabular}%
  \label{tab:Spearman}%
\end{table}%

\subsubsection{Performance comparison} % PERFORMANCE MEAN COMPARISON DEPENDING ON THE PATTERNS

The software delivery performance that each pattern shows was compared to analyze the significant differences between them.  The Wilcoxon signed-rank test was run to compare the patterns in pairs, obtaining significant differences between patterns A and D for the indicator LT ($Z=2.385$; $\textrm{p-value}=0.017$). Significant differences were also obtained between patterns B and D for LT ($Z=-2.133$; $\textrm{p-value}= 0.033$) and MTTR ($Z=-2.384$; $\textrm{p-value}=0.017$). This indicates that the companies that followed Pattern D (i.e., Full cross-functional DevOps team) perform better than less mature patterns such as patterns A and B. On the contrary, for the indicator DF we could not reject the null hypothesis (equality of means), since the p-value exceeds $0.05$. In this regard, the frequency of deployment has little to do with the pattern as explained before, since companies with less mature patterns usually deploy once per sprint as many companies with more mature  patterns do.

%Iba a añadir una línea para justificar el siguiente analisis pero a lo mejor se volvía en nuestra contra :S
%Considering that from the point of view of quantitative analysis the number of companies is not too high, ... 
%If there were more organizations in the study, other significant differences would probably have been observed...

In order to deepen this issue, two groups of organizations were made: i) a group, termed A-B, was created bringing together the organizations that followed the patterns A and B; ii) a group, termed C-D, was created bringing together the organizations that followed the patterns C and D. The software delivery performance obtained by each group was compared and the significant differences between them were analyzed. Table~\ref{tab:ranges2groups} displays the average range, together with the samples included in each group. Statistically significant differences were found in overall performance indicators, as shown in Table~\ref{tab:stat2groups}. In all the cases the null hypothesis (equality) is rejected at confidence level $0.05$. Therefore, we conclude that \textbf{companies that implement the most mature team patterns (i.e., Patterns C and D) achieve better software delivery performance indicators}.

\begin{table}[htbp]
  \centering
  \caption{Ranges for two groups of organizations}
    \begin{tabular}{p{5.355em}ccc}
    \hline
    
          & Group of patterns & $N$ & Avg.\ range \\
    \hline
    \multirow{2}{*}{LT} & \multicolumn{1}{c}{A-B} & 12    & \multicolumn{1}{r}{20.92} \\
          & \multicolumn{1}{c}{C-D} & 18    & \multicolumn{1}{r}{11.89} \\\hline
    %     Total &  & 30    &   \\\hline
    \multirow{2}{*}{MTTR}  & \multicolumn{1}{c}{A-B} & 12    & \multicolumn{1}{r}{19.83} \\
          & \multicolumn{1}{c}{C-D} & 18    & \multicolumn{1}{r}{12.61} \\\hline
    %     Total &  & 30    &   \\\hline
    \multirow{2}{*}{DF} & \multicolumn{1}{c}{A-B} & 12    & \multicolumn{1}{r}{11.33} \\
          & \multicolumn{1}{c}{C-D} & 18    & \multicolumn{1}{r}{18.28} \\
    %     Total & & 30    &   \\
    \hline
    \end{tabular}%
  \label{tab:ranges2groups}%
\end{table}%

\begin{table}[htbp]
  \centering
  \caption{Test statistics for two groups of organizations}
    \begin{tabular}{lrrr}
    \hline
          & {LT} & {MTTR} & {DF} \\
    \hline  \hline
    $U$ of Mann-Whitney & 43    & 56    & 58 \\
    $W$ of Wilcoxon & 214   & 227   & 136 \\
    $Z$     & -2.778 & -2.249 & -2.142 \\\hline
    p-value  & 0.005 & 0.025 & 0.032 \\
    %Asymp.\ Sig.  & 0.005 & 0.025 & 0.032 \\
    \hline
    \end{tabular}
  \label{tab:stat2groups}
\end{table}

\subsection{Quantitative assessment of the taxonomy}

This section assess whether the taxonomy we built through qualitative analysis (see Table \ref{tab:patterns}) is confirmed from a quantitative perspective. To that end, a K-means clustering analysis focused on the key variables shown in Table~\ref{tab:variables} (i.e., leadership from management, shared ownership, collaboration frequency, organizational and cultural silos, and autonomy) was carried out. Then, we compared the groups obtained from the clustering with the patterns we identified in Table~\ref{tab:patterns} and we analyzed their profile according to the final centers.

First, a K-means analysis with two clusters (see Table~\ref{tab:centers2}) reveals two groups of organizations that are practically the same as the patterns previously termed as A-B and C-D. Cluster 1 includes 11 organizations and Cluster 2 consists of 19.  The specific companies that make up each cluster are shown in Table~\ref{tab:correspondance}(a), together with the pattern previously assigned (pattern from A to D). Table~\ref{tab:centers2} shows the final centroids of every variable, so that the profile of every cluster is drawn. The comparison of software delivery performance between the resulting clusters also revealed similar results to those obtained by patterns A-B and C-D. In particular, the Wilcoxon test test conducted on the companies automatically grouped into the two clusters evidences significant differences in the indicators LT ($Z= -2.911$; $\textrm{p-value} = 0.004$) and DF ($Z= -2.396$; $\textrm{p-value} = 0.017$). No significant differences were obtained for the variable MTTR ($\textrm{p-value} = 0.077$).

\begin{table}[htbp]
  \centering
  \caption{Final centroids for two clusters}
    \begin{tabular}{|p{3.2cm}|l|l|}
    \hline
    & \multicolumn{2}{c|}{\textbf{Clusters}} \\\cline{2-3}
    & $N=11$ & $N=19$ \\
    \hline
    Leadership & Multiple & Single \\
    Shared Ownership & Crawl & Run \\
    Collaboration Frequency & Weekly & Daily \\
    Organizational Silos & Yes & No \\
    Cultural Silos & Yes & No \\
    Autonomy & Low & Medium \\
    \hline
    \end{tabular}%
  \label{tab:centers2}%
\end{table}%

Second, a K-means analysis with four clusters (see Table~\ref{tab:centers4}) reveals four groups that are very similar to the four team structure patterns obtained through the qualitative analysis (i.e. Patterns A, B, C, and D). The specific companies that make up each of the four clusters are shown in Table~\ref{tab:correspondance}(b), together with the pattern previously assigned. The centroid values of the key variables in each group are shown in Table~\ref{tab:centers4}. These results are practically the same as the values assigned to the variables for each pattern and the organizations also coincide to a great extent. The match is not exact because i) in the organizational patterns we allowed the variables to adopt two close values, but that is not possible in a K-means analysis; ii) some companies seemed to be transitioning between different patterns; iii) the line separating certain related patterns when a company is transitioning is not very thick (e.g., pattern C and D). 

\begin{table}[htbp]
  \centering
  \scriptsize
  \caption{Final centroids for four clusters}
    \begin{tabular}{|p{2.7cm}|l|l|l|l|}
    \hline
    & \multicolumn{4}{c|}{\textbf{Clusters}} \\[0.07cm]\cline{2-5}
    %& 1 & 2 & 3 & 4 \\
    \vspace{0.2cm}& $N=5$   & $N=3$   & $N=10$  & $N=12$ \\
    \hline
 
    Leadership & Single   & Single   & Multiple    & Single \\
    Shared Ownership & Run   & Walk  & Crawl & Run \\
    Collaboration Frequency & Daily & Weekly & Weekly & Daily \\
    Organizational Silos & No    & No    & Yes   & No \\
    Cultural Silos & Vestigial & Vestigial & Yes   & No \\
    Autonomy & Medium & Medium & Low   & Medium \\
    \hline
    \end{tabular}%
  \label{tab:centers4}%
\end{table}%

\renewcommand{\arraystretch}{1.1}

\begin{table*}[htbp]
  \centering
  \caption{Correspondence between patterns and clusters: (a) for 2 clusters, (b) for 4 clusters)}
    %\begin{tabular}{rrrrrrrrrrrr}
    \begin{tabular}{|lc|lc||lc|lc|lc|lc|}
    \hline
    \multicolumn{4}{|c||}{\normalsize\textbf{(a) K-means 2 clusters}} & \multicolumn{8}{c|}{\normalsize\textbf{(b) K-means 4 clusters}} \\    \hline\hline
   \multicolumn{2}{|c|}{Cluster 1} & \multicolumn{2}{c||}{Cluster 2} & \multicolumn{2}{c|}{Cluster 1} & \multicolumn{2}{c|}{Cluster 2} & \multicolumn{2}{c|}{Cluster 3} & \multicolumn{2}{c|}{Cluster 4} \\
   Organ. & {Pattern} & Organ.   & {Pattern} & {Organ.} & {Pattern} & {Organ.} & {Pattern} & {Organ.} & {Pattern} & {Organ.} & {Pattern}  \\
    \hline
   ID19 & B     & ID8   & D     & {ID8 } & D     & {ID2 } & C     & {ID1 } & A     & {ID3 } & D \\
   ID1 & A     & ID18  & D     & {ID18} & D     & {ID19} & B     & {ID4 } & A     & {ID9 } & D \\
   ID4 & A     & ID24  & C     & {ID24} & C     & {ID27} & B     & {ID5 } & A     & {ID10} & D \\
   ID5 & A     & ID11  & C     & {ID11} & C     &       &       & {ID6 } & B     & {ID13} & C \\
   ID6 & B     & ID17  & C     & {ID17} & C     &       &       & {ID7 } & B     & {ID15} & D \\
   ID7 & B     & ID2   & C     &       &       &       &       & {ID12} & A     & {ID22} & D \\
   ID12 & A     & ID27  & B     &       &       &       &       & {ID16} & A     & {ID23} & D \\
   ID16 & A     & ID3   & D     &       &       &       &       & {ID20} & A     & {ID26} & D \\
   ID20 & A     & ID9   & D     &       &       &       &       & {ID25} & A     & {ID28} & D \\
   ID25 & A     & ID10  & D     &       &       &       &       & {ID14} & B     & {ID29} & D \\
   ID14 & B     & ID13  & C     &       &       &       &       &       &       & {ID30} & C \\
          &       & ID15  & D     &       &       &       &       &       &       & {ID31} & D \\
          &       & ID22  & D     &       &       &       &       &       &       &       &  \\
          &       & ID23  & D     &       &       &       &       &       &       &       &  \\
          &       & ID26  & D     &       &       &       &       &       &       &       &  \\
          &       & ID28  & D     &       &       &       &       &       &       &       &  \\
          &       & ID29  & D     &       &       &       &       &       &       &       &  \\
          &       & ID30  & C     &       &       &       &       &       &       &       &  \\
          &       & ID31  & D     &       &       &       &       &       &       &       &  \\    \hline
    \end{tabular}%
  \label{tab:correspondance}%
\end{table*}%

The comparison of software delivery performance between the resulting clusters revealed similar results to those obtained by patterns A, B, C, and D. Really, we can assert that differences increase with respect to the differences between patterns. The most significant differences were obtained between Cluster 3 (which mainly represents pattern A) and Cluster 4 (which mainly represents pattern D). In particular, Wilcoxon test provided significant differences in the three indicators: LT ($Z=-3.023$ and $\textrm{p-value} = 0.003$), MTTR ($Z=-2.110$ and $\textrm{p-value}=0.036$), and FD ($Z=-2.405$ and $\textrm{p-value}=0.016$). Moreover, significant differences were obtained for variable LT between Cluster 2 (which mainly represents pattern B) and Cluster 4 (pattern D), with $Z= -2.122$ and $\textrm{p-value} = 0.034$.

In summary, this analysis shows that \textbf{clusters reported by the K-means  are very similar to the taxonomy we described in Section~\ref{sec:model}} and confirms that \textbf{the most mature team patterns (i.e., Patterns C and D) obtain the best software delivery performance indicators}. %This coincides with the insights gained during the qualitative analysis, which also point out this implication of the organizational patterns.

%%%%%%%%%%%%%%%%%%%%%%%%%%%%%%%%%%%%%%%%%%%%%%%%%%%%%%%%%%%%%%%%%%%%%%%%%%%%%%%%%%%%%%%%%%%%%%%%%%%%
%%%%%%%%%%%%%%%%%%%%%%%%%%%%%%%%%%%%%%%%%%%%%%%%%%%%%%%%%%%%%%%%%%%%%%%%%%%%%%%%%%%%%%%%%%%%%%%%%%%%
\section{Threats to Validity and Reliability and Limitations}
\label{sec:validity}
%Es necesario identificar y considerar las amenazas a la validez de un estudio para juzgar la calidad del mismo, especialmente en estudios empíricos, ya sean cuantitativos o cualitativos [Feldt, 2010]. 

This section describes how we aimed for quality criteria, both validity and reliability of qualitative studies \cite{Creswell:2017}. Specifically we aimed for \textit{credibility} (also referred to as \textit{trustworthiness}), \textit{resonance}, \textit{usefulness}, \textit{transferability}, \textit{dependability}, and \textit{conformability} \cite{charmaz:2014,Lincoln:1985,ralph:2020}. To mitigate threats to the validity and the reliability, we applied the strategies described in Section~\ref{subsec:validityprocedures} as follows.

\subsection{Threats to qualitative validity}

To mitigate threats to qualitative validity we checked for the accuracy, trustworthiness, and \textit{credibility} of the findings. The strategies implemented for this purpose were (1) triangulation, (2) spending prolonged time in the field, and (3) external auditory procedures. (1) We conducted two types of triangulation as follows. Data triangulation so that we interviewed 46 stakeholders from 31 companies, which means we collected data in different times and locations, and from different populations. Methodological triangulation so that we used different methods to collect data, i.e., interviews, workshops, observations, and retrieval of a set of software delivery performance indicators. On the one hand, the GT study mainly focused on the analysis of the interviews while observations and workshop annotations were recorded in a research diary and used to triangulate and complete data omissions. On the other hand, we perform an statistical analysis focused on the resulting patterns of the GT study and the indicators we retrieved. (2) We spent more than 3 years in retrieving and analyzing all these data. (3) Finally, preliminary results were previously reported and presented in an international conference \cite{Diaz:2019}, which can be considered as an external auditory. 

Qualitative validity is also concern to \textit{resonance}. A key strategy to that end is member checking, so that the participants first received the interview transcription, and then, some of them received preliminary results to ensure the correctness of our findings through a presentation in an industrial event\footnote{\url{https://bit.ly/2LDRA2W}, last accessed January 2020.}.

Finally, qualitative validity is also concerned to \textit{transferability} so that the data were iteratively gathered from a number of companies that is large enough to build a complete picture of the phenomenon. In this study, 31 companies were included, and 46 stakeholders participated in the interviews. This multiplicity is what provides the basis for \textit{analytical generalization} or \textit{theoretical generalization}, where the results are extended to cases that have common characteristics and hence for which the findings are relevant \cite{Wohlin:2012}.

\subsection{Threats to qualitative reliability}

To mitigate threats to qualitative reliability we checked for the consistency and \textit{dependability} of procedures. The strategies implemented for this purpose were (1) a rich description of the research, (2) report of discrepant information, and (3) bias clarification. (1) We documented as many of the steps of the procedures as possible and provided a full description of involved organizations and teams as confidentiality and ethical issues allowed. (2) We documented some memos to highlight some minor contradictions. (3) We conducted a double-check of transcripts: a first one to eliminate possible errors in literal transcription, as quite half of the interviews were transcribed by students and they were not familiar with certain terminology; and a second one to eliminate different interpretations of a same concept. For example, related to the question about autonomy, ID2 says that they are not an autonomous teams because they have dependencies to deploy into production (a check point from business), but ID23 says that they are autonomous, although reviewing other questions, they later say that they have dependencies to deploy into production. Although this was the most time-consuming task, the following reliability procedures are also undoubtedly the most important: (i) identify possible drifts  or shifts in the definition and meaning the codes, which were mainly addressed and documented by memos as the described in Section~\ref{sec:theory}, and (ii) coordination and communication among coders through online meetings through Microsoft Teams, which were documented and mostly of them recorded. 

Qualitative reliability is also concern to conformability. In this regard, Section~\ref{sec:theory} presents clear chain of evidence from interviewee quotations to the proposed categories concepts.

\subsection{Limitations}
Finally, some limitations concern to the lack of inter-rater agreement analysis to improve reliability of coding. Although, the authors of this article have most often applied inter-coder agreement within thematic analysis, and found it improved the quality, transparency, and reliability, we also found that the recursive and incremental nature of grounded theory’s analytic process makes inter-coder agreement difficult to implement. Thus, GT does not guarantee that different researchers working with the same data would code identical categories and achieve identical theory, as depends of the the researchers' theoretical sensitivity, which is consistent with our epistemological philosophy.
%our theory can always be expanded and refined with new data.  %Similarly, according to Glaser and Strauss, judging theoretical saturation is never precise and depends on the researchers' theoretical sensitivity [10]. Also, a theory can always be expanded and refined with new data.

%%%%%%%%%%%%%%%%%%%%%%%%%%%%%%%%%%%%%%%%%%%%%%%%%%%%%%%%%%%%%%%%%%%%%%%%%%%%%%%%%%%%%%%%%%%%%%%%%%%%
%%%%%%%%%%%%%%%%%%%%%%%%%%%%%%%%%%%%%%%%%%%%%%%%%%%%%%%%%%%%%%%%%%%%%%%%%%%%%%%%%%%%%%%%%%%%%%%%%%%%
\section{Related Work}
\label{sec:relatedwork}

Chronologically, although the study by Iden et al. \cite{Iden:2011} did not explicitly mention DevOps, we can consider that is one of the seminar papers that empirically analyzed the conflict between development and operations teams when they have to collaborate. This follows the idea initiated by P. Debois \cite{Debois:2008} and Flickr employees \cite{Allspaw:2009}.  Iden et al. \cite{Iden:2011} used the Delphi method (brainstorming with 42 Norwegian IT experts, reduction and ranking) to provide a key baseline for analyzing the problems that reveal the lack of cooperation between developers and IT operations personnel. Later, Nybom et al. \cite{nybom:2016} interviewed 14 employees of an organization to analyze the benefits of dev \& ops collaboration, such as improved trust and smoother work flow, and costs, such as new sources for friction among the employees and risk for holistically sub-optimal service configurations. 

Some years later, Erich  et al. \cite{Erich:2014} performed a systematic mapping study to analyze the benefits of this cooperation between development and operations, and Smeds et al. \cite{Smeds:2015} interviewed 13 subjects in a software company adopting DevOps to research the main defining characteristics of DevOps and the perceived impediments to adopting DevOps. Later, Lwakatare et al. \cite{Lwakatare:2016b} used multi-vocal literature and three interviews from one case company to describe what DevOps is and outline DevOps practices according to software practitioners. In the same year, Riungu-Kalliosaari et al. \cite{Lwakatare:2016a} conducted a qualitative multiple-case study and interviewed the representatives of three software development organizations in Finland to answer how industry practitioners perceive the benefits of DevOps practices in their organization and how they perceive the adoption challenges related to DevOps. 

In the following years a greater number of studies conducted empirical research by involving an increasing number of companies. Luz et al. \cite{LuzPinto:2019} conducted grounded theory analysis on about 15 scenarios of successful DevOps adoption in companies to build a model. Recently, Leite et al. \cite{Leite:2020} published an exhaustive survey on DevOps concepts and challenges, in which through a method inspired by systematic literature review and grounded theory, they analyze practical implications for engineers, managers and researchers.

However, there exist few literature with clear focus on analyzing team structures. One of first papers in analyzing team structures is the work by Shahin et al. \cite{shahin:2017}. They conducted a mixed-method empirical study, which collected data from 21 interviews in 19 organizations and a survey with 93 practitioners. They also identified four common types of team structures: separate Dev and Ops teams with higher collaboration; separate Dev and Ops teams with facilitator(s) in the middle; small Ops team with more responsibilities for Dev team; and no visible Ops team. Soon after, we already started to sketch a first version of the taxonomy presented in this paper and started to correlate the DevOps organizational patterns with software delivery performance \cite{Diaz:2019}. Shortly afterwards, Leite et al. \cite{LeitePinto:2020} \cite{LeitePinto2:2020} collected data from 27 IT professionals and identified also four patterns: siloed departments, classical DevOps, cross-functional teams, and platform teams. 

According to our constructivist stance, we believe that there are several realities that can be discovered from different approaches. Although these studies \cite{shahin:2017} \cite{LeitePinto:2020} \cite{LeitePinto2:2020} and the present work have come up with some common elements, such as organizational structures based on development and operations silos, or the existence of cross-functional product teams, each one brings different insights. For example, we found that in every organizational pattern there is always a horizontal DevOps team, which can be formed by people exclusively dedicated to platforms (i.e., DevOps platform teams, DevOps CoE) or by people from each product team specialists in DevOps (i.e., DevOps chapter), that provides some service to the product teams, which is essential to decrease the cognitive load of these product teams and improve their productivity. In addition, each study reaches certain structures through different observations and variables. For example, we observed a set of variables, such as leadership from management or shared ownership, which from our perspective are essential to understand the organizational structure of teams. However, other studies had not paid so much attention to these variables as we do. Furthermore, each study uses different research methods. For example, we carried out, in addition to a qualitative study, an in-depth statistical analysis to (i) check whether certain patterns imply better performance, and (ii) assess whether clustering identified the same patterns we qualitatively had identified. In this way, all the differentiating elements of each study complement each other, thus allowing to reach a more complete and solid theory. It is important to take into account that the mentioned researches are contemporary and have been carried out in different parts of the world, so that a triangulation of studies could be later analyzed in a systematic study.

Finally, Bahadori \& Vardanega \cite{Bahadori} analyzed the importance of platform teams. Specifically, they discussed why product teams require infrastructure agility and how dynamic orchestration of infrastructure delivery (e.g., in Cloud environment) may accelerate software delivery. They stated that the Cloud, containers and microservices, dynamically orchestrated (e.g., using technologies such as Kubernetes, Openshift, etc.), enable an effective team structure so that developers and operators use Infrastructure-as-a-Service as its platform. Thus, this paper showed, through an experiment, the importance of horizontal DevOps teams (either platform teams or CoEs) by improving a set of indicators, such as system response time. This study is complementary to ours since we show the importance of platform teams through a grounded theory study that involves real companies and later statistical analysis based on software delivery performance.

%esta tabla la veo obviable
%\begin{table}[!h]
%\centering
%\small
%\caption{Research questions addressed in related studies}
%\label{tab:relatedwork}
%\begin{tabular}{p{5cm} p{3cm}}
%& \\
%\hline
%\textbf{Research Question} & \textbf{References}\\ \hline 
%Problems why companies move to DevOps & \cite{Senapathi:2018} \cite{LuzPinto:2019} [EMP]\\
%DevOps concept and characteristics &  \cite{Smeds:2015} \cite{Lwakatare:2016b}  \cite{Erich:2017} \\
%DevOps expectations - perception of benefits & \cite{LuzPinto:2019} [EMP]\\
%DevOps enablers - supportive factors &  \cite{Erich:2017} \cite{Senapathi:2018} \\
%DevOps practices - how is DevOps implemented & \cite{Lwakatare:2016b} \cite{Erich:2017} \cite{dora:2018} \cite{puppet:2018}  \\
%DevOps benefits - effects &  \cite{Erich:2014} \cite{Lwakatare:2016a} \cite{Erich:2017} \cite{Senapathi:2018} \cite{dora:2018} \\
%DevOps challenges - impediments  & \cite{Smeds:2015} \cite{Lwakatare:2016a} \cite{Senapathi:2018} \cite{Kuusinen:2018}   \\
%Security concerns in DevOps adoption & \cite{Saima:2020} \\
%Team Topologies &  \cite{LeitePinto:2020}\\
%\hline
%\end{tabular}
%\end{table}

%%%%%%%%%%%%%%%%%%%%%%%%%%%%%%%%%%%%%%%%%%%%%%%%%%%%%%%%%%%%%%%%%%%%%%%%%%%%%%%%%%%%%%%%%%%%%%%%%%%%
%%%%%%%%%%%%%%%%%%%%%%%%%%%%%%%%%%%%%%%%%%%%%%%%%%%%%%%%%%%%%%%%%%%%%%%%%%%%%%%%%%%%%%%%%%%%%%%%%%%%
\section{Conclusion}
\label{sec:conclusion}

This paper presents a taxonomy for DevOps team structure patterns. It reveals four team structures that companies have been adopting, from the most emerging ones based on a sporadic collaboration between Dev and Ops to the most consolidated ones, in which silos were broken. They are cross-functional and highly autonomous teams, and thus, achieve better software delivery performance. This taxonomy is the result a Grounded Theory study that analyzed 31 multinational software-intensive companies. To improve the research validity and reliability we used various strategies (e.g., data triangulation, rich description, use of external auditors, etc.) that mitigate the problems inherent to qualitative methods and reinforce our findings. This qualitative analysis was subsequently supported and reinforced with a statistical analysis. 

%quizás podríamos poner un párrafito como este pero a lo mejor podríamos refinarlo un poco
Practitioners can use the presented taxonomy to evolve the organizational structure of their company to one with which to effectively implement DevOps, as well as to evaluate their current degree of maturity regarding DevOps. Furthermore, SE researchers can use this work as a guide or example for conducting qualitative research involving Grounded Theory.

\subsection{Lessons Learnt}

%Ibamos a poner aquí lo de la autonomia (que unos decían que tenían autonomia pero veiamos que no...)? Esto nos daría pie para hablar de las dificultades de la cualitativa... aunque no recuerdo si ya estaba puesto en otro lado (o habías pensado ponerlo en otro lado)

Horizontal DevOps teams work closely with product teams. We found that the more strong horizontal teams are, the more services they can offer to the product teams, which enables better software delivery performance as shown in Section \ref{sec:implications}. Therefore, if an organization wants to move forward with its DevOps adoption, it should start by creating a strong horizontal DevOps team (i.e., a DevOps Center of Excellence), which provides an added value beyond the service of the CI/CD platforms. For example, actions such as evangelization, mentoring or the temporarily human resources assignment, help product teams to acquire DevOps practices and to establish the DevOps culture. This recommendation is critical to address one of the main problems that motivated companies to adopt DevOps, i.e.,  too much time for releasing, as we analyzed in a previous research \cite{ese}.

%We confirm the words of relevant consultants such as Skelton and Pais creating “team APIs” in order to reduce cognitive load, and the benefits of building a “thinnest viable platform” Defining a team API can reduce the cognitive load on understanding the communication patterns, especially in distributed, remote working. 

Furthermore, although it is not the objective of this paper, we observed that some organizational patterns usually lead to some modern architectures, i.e., it seems that organizational team patterns may impact on  software architecture. For example, in many cases the most mature team patterns result in microservice architectures due to the valuable support of a DevOps CoE. On the contrary, the less mature patterns result in monolithic architectures, even when a microservice approach would bring great benefits. This insight is related to one of the main drivers of organizations when adopting DevOps \cite{ese}, i.e., the need to initiate a transformation due to technological obsolescence or large architectural, infrastructural and organizational changes.

Finally, we also observed that some patterns, such as Interdepartmental Dev \& Ops collaboration, which shows less maturity in the DevOps adoption, may becomes a silver bullet to accelerate value delivery in large companies. Some  large companies are not able to adopt a “pure” DevOps with stable  and cross-functional product teams because, although results might be better, the cost would be unbearable. Deepening in this issue, it is worth mentioning that although some organizations carry out all their projects following DevOps, most organizations take a bimodal approach and they also use the traditional approach for certain projects that do not need what DevOps offers (e.g., a decrease of time-to-market, etc.). This is because to adopt DevOps is costly and it is often smart to use DevOps for new applications that require frequent updates and maintain the traditional approach for other projects.

%No estoy convencido de si conviene meter aquí, ya casi al final, un extracto... aunque si tú lo ves claro pues adelante.
%Hence, some organizations adopt mixed models as the one described by ID08:
%\vspace{0.2cm}
%\noindent \fbox{\begin{minipage}{8.5cm}
%[ID08] \textit{“The development was organized through vertical teams (channels, invoicing, processing, mediation, client system, sales, etc.). The production level was similarly organized through horizontal teams (e.g. web, war, operating system, etc.). Then, matrix areas are generated in which you put people from operations in a development team working on demand management, usability requirements, etc. They collaborate and move the needs from development to production and from production to development. Development and operations communicate through this matrix in which a one person from business, one person from development and one person from operations make the link between silos.”}
%\end{minipage}}
%\vspace{0.2cm}

\subsection{Future Work}

Although the team structure patterns impact on the software delivery performance, as we shown in Section \ref{sec:implications}, the performance is probably also affected by many other factors, such as the tools for automating testing, CI/CD, monitoring, and so on. As future work, we plan to study how these factors impact on the performance of product teams. Moreover, it would also be interesting to explore DevOps adoption barriers, and more specifically, the problems or barriers associated with the adoption of each of the organizational team patterns defined in this research.

Finally, we observed that many organizations are investing in technical debt reduction and this seems to have a medium-term impact on the performance achieved. An in-depth research of this issue could also be of interest.

\ifCLASSOPTIONcompsoc
  \section*{Acknowledgments}
\else
  \section*{Acknowledgment}
\fi

This research project is being performed thanks to Vass, Clarive, Autentia, Ebury, Carrefour, Vilt, IBM, AtSistemas, Entelgy, Analyticalways, Mango eBusiness, Adidas, Seur, Zooplus, as well as other participating companies that prefer to remain anonymous.

\ifCLASSOPTIONcaptionsoff
  \newpage
\fi



%\bibliographystyle{IEEEtran}

\bibliography{references.bib}{}
\bibliographystyle{IEEEtran}

%H.~Kopka and P.~W. Daly, \emph{A Guide to {\LaTeX}}, 3rd~ed.\hskip 1em plus
%\end{thebibliography}


%\end{IEEEbiography}


%\newpage



%\vfill


\end{document}